\documentclass[graybox, nosecnum]{svmult}

\usepackage{mathptmx}       
\usepackage{helvet}         
\usepackage{courier}        
\usepackage{type1cm}        
\usepackage{makeidx}         
\usepackage{graphicx}        
\usepackage{multicol}        
\usepackage[bottom]{footmisc}
\usepackage{hyperref}        
\usepackage{soul}            
\hypersetup{colorlinks=true,linkcolor=magenta,citecolor=blue,urlcolor=blue}
\usepackage[square,numbers,compress]{natbib}
\makeindex             

\newcommand\solphys{{Solar Physics}}%
\newcommand\apj{{Astrophys.\ J.}}%
\newcommand\apjl{{Astrophys.\ J.}}%
\newcommand\aj{{Astron.\ J.}}%
\newcommand\aap{{Astron. Astrophys.}}%
\newcommand\mnras{{Mon.\ Not.\ R.\ Astron.\ Soc.}}%
\newcommand\nat{{Nature}}%
\newcommand\araa{{Annu.\ Rev.\ Astron.\ Astrophys.}}%

\begin{document}
\title*{The HaloSat and PolarLight CubeSat Missions for X-ray Astrophysics}
\author{Hua Feng and Philip Kaaret}
\institute{Hua Feng \at Department of Astronomy, Tsinghua University, Beijing 100084, China \email{hfeng@tsinghua.edu.cn}
\and Philip Kaaret \at Department of Physics and Astronomy, University of Iowa, Iowa City, IA 52242, USA \email{philip-kaaret@uiowa.edu}}
%
%
\maketitle
\abstract{Astronomical observations in the X-ray band are subject to atmospheric attenuation and have to be performed in the space. CubeSats offer a cost effective means for space-based X-ray astrophysics but allow only limited mass and volume.  In this article, we describe two successful CubeSat-based missions, HaloSat and PolarLight, both sensitive in the keV energy range. HaloSat was a 6U CubeSat equipped with silicon drift detectors. It conducted an all-sky survey of oxygen line emission and revealed the clumpy nature of the circumgalactic medium surrounding the Milky Way. PolarLight is a dedicated X-ray polarimeter performing photoelectron tracking using a gas pixel detector in a 1U payload.  It observed the brightest X-ray objects and helped constrain their magnetic field or accretion geometry. On-orbit operation of both missions for multiple years demonstrates the capability of CubeSats as an effective astronomical platforms. The rapid time scales for development and construction of the missions makes them particularly attractive for student training.}

\section{Keywords} 
CubeSat, X-ray, detector, space astronomy, high energy astrophysics, Milky Way halo, polarimetry

\section{Introduction}

CubeSats are a class of small satellites that follow certain design specifications. A CubeSat may occupy a volume of one, several, or a fraction of a `unit' (U) defined with dimensions of 10~cm $\times$ 10~cm $\times$ 10~cm.  Most CubeSats have a size of 1U, 3U, 6U, or 12U. Use of fixed design specifications and interfaces enables CubeSats to fly on a variety of launch vehicles and decouples construction of the CubeSat from the specific launch opportunity. This allows CubeSats to piggyback on the launch of large missions, greatly reducing cost. CubeSats have become an effective and affordable means to demonstrate new technologies for space science and to train students in the development of space missions.

Along with the growing popularity of CubeSats, there has been increasing interest in their use for space-based astronomy \cite{Shkolnik2018}.  Unlike balloons that fly within the atmosphere, CubeSats operate in space, enabling observations in wavebands subject to substantial atmospheric attenuation such as the UV and soft X-ray bands. Compared with the very short durations of sounding rockets flights, CubeSats can operate for years. Astrophysical missions based on CubeSats may take advantage of long duration observations to compensate for the small telescope apertures and detection areas that result from the constraints on CubeSat mass and volume. Due to their low cost, CubeSat missions can be designed for very specific scientific goals.  This enables, e.g., observing programs devoted to deep observations of a limited number of targets enabling long-term monitoring. 

In this chapter, we focus on CubeSat missions for extrasolar X-ray astrophysics, in particular on two missions launched in recent years: HaloSat \cite{Kaaret2019} and PolarLight \cite{Feng2019}. Both are sensitive in the energy range of a few keV. Gamma-ray CubeSat missions are described and discussed in Section IV of this book. There have also been several solar X-ray missions based on CubeSats \cite{Moore2018,Mason2020}; they are not included in this chapter.  HaloSat performed an all-sky survey of oxygen line emission in the soft X-ray band to attempt to better understand the distribution of hot plasma in the Milky Way halo \cite{Kaaret2020}. PolarLight is a collimated X-ray polarimeter observing the brightest X-ray sources in the sky in order to diagnose their magnetic field or accretion geometry \cite{Feng2019}.  In the following sections, we introduce the instrumentation, operation, and performance of the two missions, and briefly summarize the science results. Then, we discuss a few lessons learned in building CubeSats for space astronomy, and how such missions can benefit student training.

\section{HaloSat}
\label{sec:halosat}

HaloSat was the first competitive-selected CubeSat funded by NASA's Astrophysics Division. The scientific goal of HaloSat was to map the spatial distribution of hot gas in the halo or circumgalactic medium of the Milky Way by mapping soft X-ray line emission \cite{Kaaret2019}. The mission was developed in less than 2.5 years from the start of funding to launch. The mission was deployed from the International Space Station on 2018 July 13, began science operations in 2018 October, and re-entered Earth's atmosphere on 2021 January 4. The data from the full mission have been archived at NASA's High Energy Astrophysics Science Archive Research Center (HEASARC) and are publicly available in the standard data formats and can be reduced and analyzed using standard tools for X-ray data analysis. This enables their use by the astrophysics community.

\subsection{Scientific Goals}

In 1956, Spitzer first suggested the existence of a `Galactic corona' or halo of hot ($10^6$~K) gas surrounding in the Milky Way to help explain of presence of clouds of gas far from the Galactic plane \cite{Spitzer1956}. The halo or circumgalactic medium (CGM) has played a key role in evolution of the Milky Way. It is the source of material for star formation and a repository for metals thereby produced \cite{Putman2012,Tumlinson2017}. The CGM may also be a repository for baryons seen in the early universe, but undetected locally\cite{Shull2012}.

The CGM is multiphase with the majority of the gas at temperatures near $10^6$~K \cite{BlandHawthorn2016}. Gas at this temperature is best studied in the X-ray band, either via absorption line or emission line spectroscopy. The primary lines used are those of oxygen, which has the highest cosmological abundance after hydrogen and helium and is not fully ionized at the $\sim 10^6$~K temperatures of the halo. Absorption line spectroscopy requires a bright background source, usually an X-ray bright active galactic nucleus (AGN), in order to obtained the high signal to noise ratio required. Given the capabilities of current X-ray observatories, only about 30 lines of sight can be probed in absorption \cite{BlandHawthorn2016}. 

X-ray emission is produced directly by the halo gas and provide a means to study the entire halo. Major X-ray observatories, mainly XMM-Newton and Suzaku, have been used to measure X-ray emission from the halo, but they were not designed for the study of diffuse emission on large angular scales. Also, the spectra obtained are often contaminated by foreground heliospheric emission that limits the accuracy of measurements of the emission from the CGM. The scientific goal of HaloSat was to map the spatial distribution of hot gas in the circumgalactic medium of the Milky Way by mapping oxygen line emission in the soft X-ray band.

\begin{figure}[t]
\centering
\includegraphics[width=0.7\columnwidth]{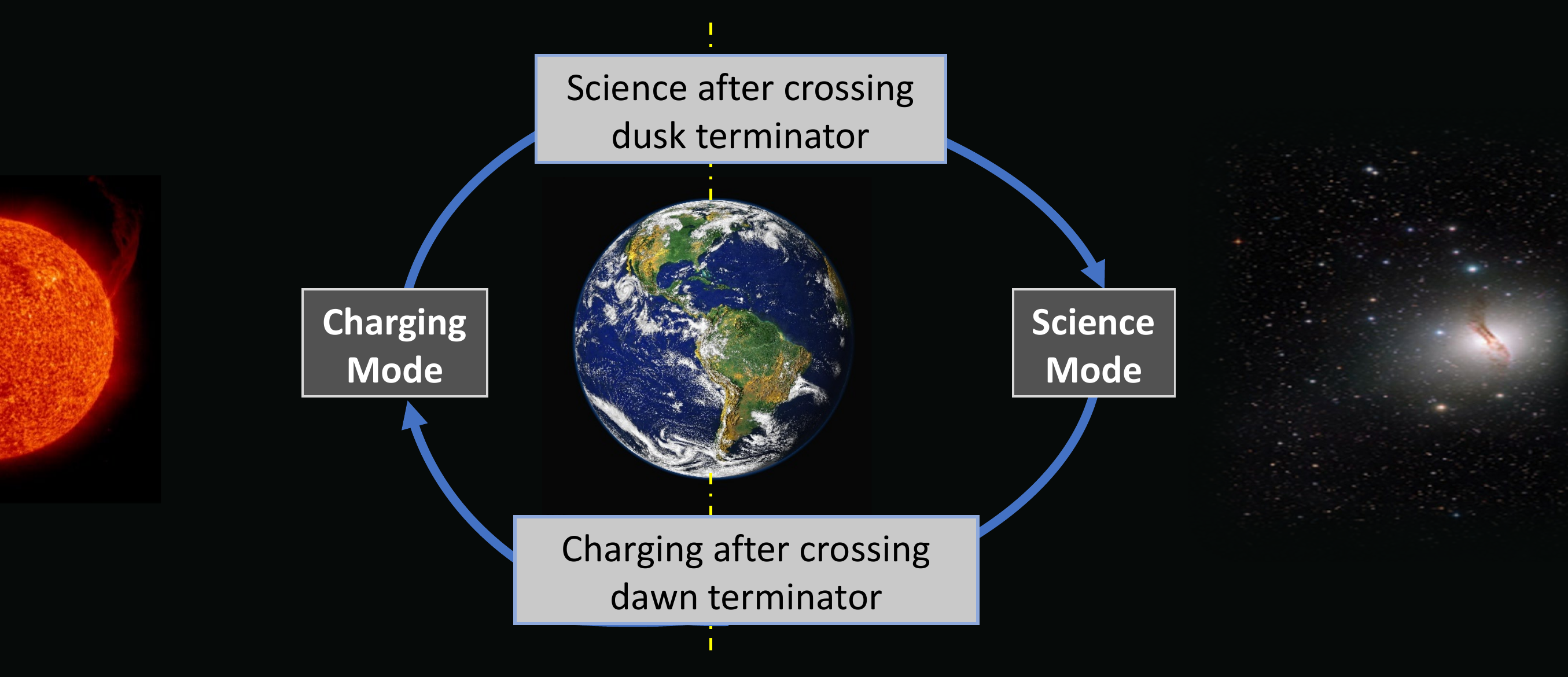} 
\caption{HaloSat operations for each spacecraft orbit. During the day half of each orbit, HaloSat oriented the solar array to charge the battery and the science instruments were switched off. During the night half of each orbit, the science instruments were operated and HaloSat pointed at two different targets for one-quarter orbit each.}
\label{fig:hs_ops}
\end{figure}

\subsection{Mission and Operations Design}

HaloSat was designed to provide good sensitivity to diffuse X-ray emission extended on large angular scales (tens of degrees) in the soft X-ray band. The mission and operations were designed to produce an all-sky survey while minimizing foreground heliospheric and magnetospheric emission.

The figure of merit for the study of diffuse emission is `grasp' which is the product of the effective area of the instrument multiplied by the field of view. HaloSat has a small effective area, about 5.1~mm$^2$ at 600~eV for each detector which is limited by the physical constraints of the CubeSat format, but has a large field of view of approximately 100 square degrees.  Three essentially identical X-ray detectors were used giving a total grasp of 17.6~cm$^2$~deg$^2$ at 600~eV. This exceeds the grasp of Chandra at launch and is within a factor of 4 of the grasp of XMM-Newton. This shows that small X-ray astronomy missions can be competitive with major missions when designed for a specific scientific goal.

Charge exchange interactions between energetic charged particles in the solar wind and neutral atoms within the solar system can leave oxygen ions in excited states that decay with the emission of an X-ray indistinguishable from X-rays produced in the halo. This is foreground emission often limits the accuracy of soft X-ray measurements of the halo. The observing strategy of HaloSat was designed to minimize the foreground due to solar wind charge exchange (SWCX; for a review see \cite{Kuntz2019}). HaloSat observations are scheduled to occur preferentially towards the anti-Sun direction. This limits the path length both through the heliosphere and through the X-ray bright parts of the magnetosphere. This provides a significant decrease in SWCX contamination, particularly relative to spacecraft such as XMM-Newton which use a fixed solar array and therefore require observation at angles roughly perpendicular to the Sun \cite{Kaaret2020}. 

Figure~\ref{fig:hs_ops} shows the sequence of operations during each HaloSat spacecraft orbit. HaloSat was in a low Earth orbit of about 90 minutes. During the day (Sun lit) half of each orbit, HaloSat orients so the solar array is normal to the Sun to maximize power received and the science instruments are switched off.  This maintained a positive power budget throughout the mission. Also, the frequent power cycling of the instrument ensured that any single event upsets or software failures within the instrument impacted less than 90 minutes of operations. During the night half of each orbit, HaloSat pointed at two different targets, one in the quarter of the orbit between the dusk terminator and spacecraft midnight and the other between midnight and the dawn terminator. This enabled HaloSat to observe towards the anti-Sun direction without Earth obscuration. The rapid changes in aspect, three per 90 minute spacecraft orbit, were feasible because the compact size and low mass of the CubeSat led to a low moment of inertia.

Instrumental background due to interactions of energetic charged particles with the detectors and the spacecraft was the other major source of background on HaloSat. Each detector was equipped with a passive shield to decrease the background as described in the following section. However, high background counting rates were encountered while passing through the South Atlantic Anomaly (SAA) or through the `polar horns' at high equatorial latitudes. The most common launch opportunities for CubeSats provided by NASA are via launches to the International Space Station (ISS). CubeSats deployed into the ISS orbit have a relatively high inclination (51.6$^{\circ}$) and spend significant time in high radiation regions at high latitudes. Ride share missions to lower inclinations could provide a significantly better radiation environment for future X-ray CubeSats.

HaloSat was built around a commercial CubeSat bus built by Blue Canyon Technologies (BCT). Use of a commercial bus provided reduced cost and improved heritage relative to a custom-designed bus. Spacecraft operations were performed by BCT based on science operations sequences planned by the University of Iowa. This provided an effective division of labor and utilized the expertise of each organization in a cost effective manner.

\begin{figure}[t]
\centering
\includegraphics[width=0.7\columnwidth]{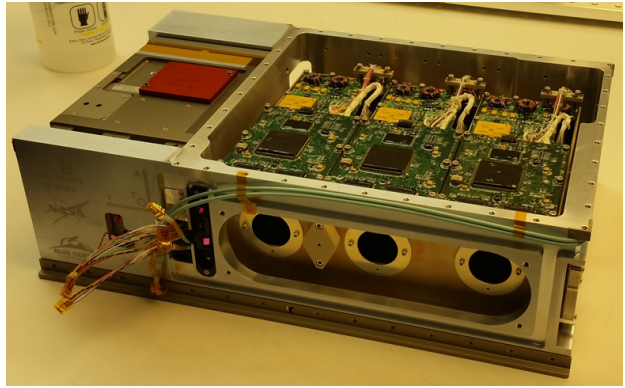} 
\caption{HaloSat after integration of the science instrument with the spacecraft. The BCT avionics module is on the left and the science instrument is on the right (there was a cover over the instrument on-orbit). The spacecraft structure has a large opening through which the three co-alligned science instrument view the sky. The fields of view were defined by the three large aluminum washers. The diamond-shaped plastic part covers an alignment mirror. Each instrument had identical and independent electronics visible on top.}
\label{fig:hs_integrated}
\end{figure}

\subsection{Science instrument development and calibration}

The science instrument and its integration with the spacecraft, see Figure~\ref{fig:hs_integrated}, is described in \cite{Kaaret2018}. Design and construction of the instrument is described in \cite{LaRocca2020} and ground calibration in \cite{Zajczyk2020}. Here, we summarize a few key features and design considerations.

HaloSat used commercially available silicon drift detectors (SDDs) from Amptek, Inc. Each SDD is a sealed package containing an X-ray sensor with an active area of 17~mm$^2$, a Si$_3$N$_4$ window, a multilayer collimator, and a thermoelectric cooler. The SDD was cooled to $-30^{\circ}$C during operations and provided an energy resolution of $\sim$85~eV (FWHM) at 677~eV \cite{Kaaret2018}. The rapid timescale of 2.5 years from start to flight was possible only because no significant technology development was needed for the mission. 

Each SDD was mounted inside a shell of 1.2~mm thick copper-tungsten composite, the mix of high and low atomic number materials acted as a graded shield. The shield was electroplated with nickel and an outer layer of gold, chosen because it has no florescence lines in the energy band of interest. The charge pulses produced by X-rays detected in each SDD were processed using an analog electronics chain with a discriminator, preamplifier, shaping amplifier, and sample/hold \cite{LaRocca2020}. Prototypes of each component of the instrument electronics were designed, built, and tested beginning early in the program with multiple iterations. The {EM} was subject to environmental testing, particularly repeated thermal cycling, to identify faults and ensure that our electronics were reliable and would meet the performance requirements over the full on-orbit temperature range and over the full mission duration. This enabled us to hold the Instrument Design Review eleven months after project start with a fully operational engineering model (EM) meeting all performance requirements.

Extensive ground calibration in vacuum over the full on-orbit temperature range \cite{Zajczyk2020} was critical in achieving reliable science results. The channel-to-energy calibration used an X-ray tube that excited florescence lines of F (677 eV), Al (1488 eV), Si (1743 eV), and Cr (5412 eV). The flight model (FM) of the HaloSat science instrument was mounted in a thermal-vacuum chamber and cycled over the temperature range from $-25^{\circ}$C to $+45^{\circ}$C. The channel-to-energy calibration was found to be temperature dependent. The calibration data were used to derive empirical relations used to correct the energy calibration on-orbit. HaloSat experienced relatively rapid temperature changes due to its low Earth orbit and small mass. Hence, the calibrations were applied on an event-by-event basis using a temperature reading obtained within 16~s of the event time. The calibration data were used to estimate the electronic noise and the energy resolution as a function of energy. Spectra of radioactive $^{55}$Fe sources were used to tune the parameters of a model of the SDD response based on code for simulation of Si X-ray detectors written by Scholze and Prokop \cite{Scholze2009}. The model was originally developed for the SDDs used on the Neutron star Interior Composition Explorer (NICER) instrument \cite{Gendreau2016} and provided to the HaloSat team by Dr. Jack Steiner. The SDDs used by HaloSat are identical to the ones used by NICER except for the type of the entrance window and the dimensions of the internal collimator. The effective area of the SDDs was calculated from the properties of the Si$_3$N$_4$ window and multilayer collimator provided by HS Foils, Oy, and the SDD dead layer thickness and active depth provided by Amptek, Inc. The effective area curve was validated on-orbit using observations of the Crab nebula. The effective area and response were used to create response files compatible with the Xspec X-Ray spectral fitting package \cite{Arnaud1996}.

\begin{figure}[t]
\centering
\includegraphics[width=0.7\columnwidth]{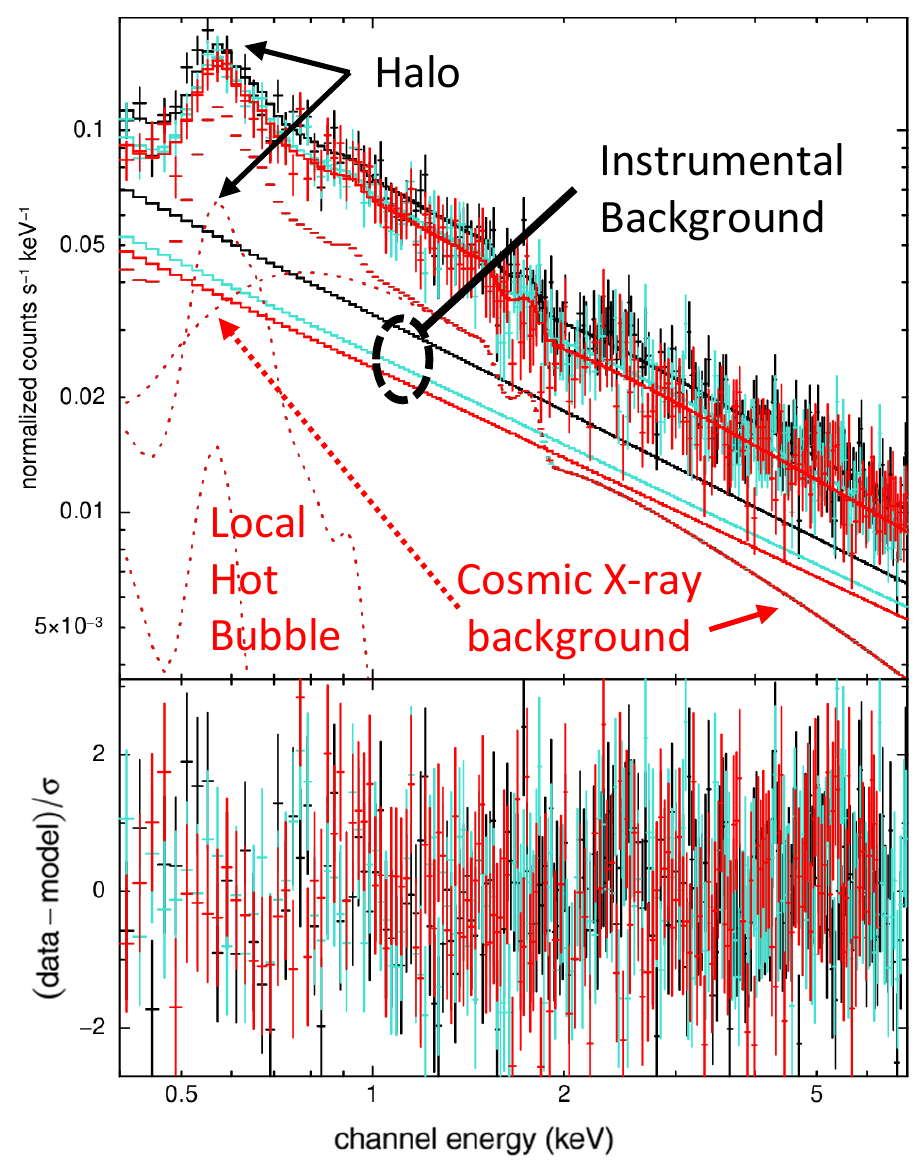} 
\caption{HaloSat spectra of a field at Galactic coordinates $l=63.91^{\circ}, b=75.93^{\circ}$. Spectra are shown for the three, co-aligned X-ray detectors. Spectra include emission from the Galactic halo, the local hot bubble, and the cosmic X-ray background as shown. These astrophysical emission components are assumed to be identical for the three detectors. The particle-induced instrumental background varies between the three detector. }
\label{fig:hs_spectrum}
\end{figure}

\subsection{Science results}

HaloSat's primary scientific goal was to constrain the spatial distribution and mass of hot gas associated with the Milky Way by mapping soft X-ray line emission from highly ionized oxygen. A study of the southern Galactic sky with HaloSat showed that the X-ray emission exhibits strong variations on angular scales of $\sim10^{\circ}$, has an overall trend versus Galactocentric radius proportional to the surface density of molecular hydrogen which is a tracer of star formation, and has a relatively small scale height of a few kpc \cite{Kaaret2020}. These results suggest that the X-ray emission is predominantly from hot plasma produced via stellar feedback. 

HaloSat conducted an all-sky survey including all Galactic latitudes. This enabled a broad range of investigations of hot plasma within the Milky Way produced by stellar and nuclear feedback. HaloSat data have been used to study the North Polar Spur that was historically considered to be the result of a nearby supernova but has recently been interpreted as related to feedback from the nuclear supermassive black hole and to the bubbles seen by the Fermi satellite at GeV energies. The HaloSat results on pressure and electron density favor a Galactic-scale event with an energy of $\sim 6\times 10^{54}$~erg and an age of $\sim$10~Myr \cite{LaRocca2020}. HaloSat data on the Cygnus superbubble, a region of bright soft X-ray emission in the direction of local spiral arm, show that the temperature and absorption are consistent across different parts, suggesting a singular origin, potentially as a hypernova remnant \cite{Bluem2020}. HaloSat observations have also been used to make the first measurement of the total X-ray luminosity of the Vela supernova remnant using CCD-level energy resolution \cite{Silich2020}.

HaloSat has also enabled studies beyond the topic of hot plasma within the Milky Way, ranging from within the solar system to clusters of galaxies. HaloSat data have been used to investigate heliospheric SWCX emission \cite{Ringuette2021}, the Large Magellanic cloud \cite{Gulick2021}, the Virgo cluster of galaxies \cite{Hewitt2021}, and to search for 3.5~keV line emission from a putative sterile neutrino produced in the decay of dark matter \cite{Silich2021}.

\begin{figure}[t]
\centering
\includegraphics[width=0.7\columnwidth]{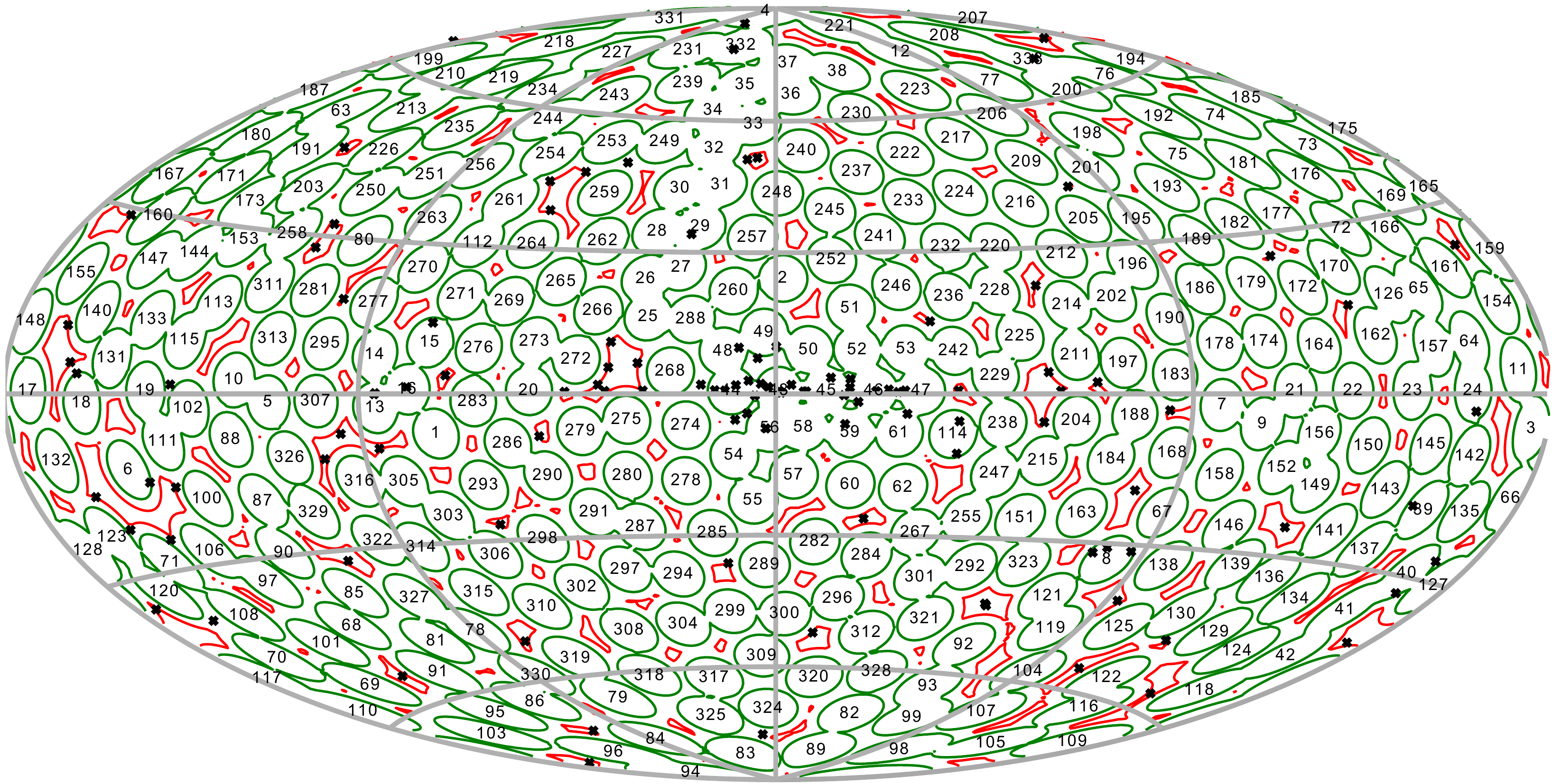} 
\caption{Map of HaloSat target fields in Galactic coordinates. The green contours shown the 10$^{\circ}$ diameter full-response field of view and the red contours show the 14$^{\circ}$ diameter zero response. The black X's mark bright X-ray sources. Figure adapted from \cite{Kaaret2018}.}
\label{fig:hs_targets}
\end{figure}

\subsection{Data archive}

The HaloSat all-sky survey tiled the sky with 334 fields with attention paid to avoid known bright X-ray sources, see Fig.~\ref{fig:hs_targets}. The data from the full HaloSat mission have been archived at NASA's High Energy Astrophysics Science Archive Research Center (HEASARC) and are publicly available for use by the astronomical community (\url{https://heasarc.gsfc.nasa.gov/docs/halosat/}). Spectral, event, and ancillary data are available and may be accessed via the standard HEASARC Browse interface or anonymous SFTP. Pre-processed spectra using standard cuts for data filtering are available for each field. All of the data files are in standard FITS (Flexible Image Transport System) format. Response matrices in standard file formats compatible with the Xspec X-Ray Spectral Fitting Package are available. This enables analysis of the data with tools commonly used throughout the X-ray astronomy community. The HaloSat HEASARC site also hosts analysis guides providing descriptions of the data, processing pipeline, data analysis, and instrumental background modeling. Public availability of archived data has been important in enhancing the scientific return of many observatories. It is important to continue this policy for CubeSats and to archive the data in standard formats and with sufficient documentation that they can be easily used by the community.

\section{PolarLight}
\label{sec:polarlight}

PolarLight is an X-ray polarimeter based on the photoelectric effect and is sensitive in the energy range of 2--8 keV.  It is a 1U payload in the CubeSat Tongchuan-1, which was launched into a LEO on 2018 October 29. As of this writing, PolarLight is still operating on orbit (for over 3 years) and is measuring X-ray polarization of the brightest X-ray sources in the sky.  The purpose of the mission is to demonstrate the new technology for future X-ray telescopes, understand the on-orbit background, and obtain science results in this field that has not been well explored. 

\subsection{Detector}

As a photoelectron is ejected by the electric field of an X-ray's electromagnetic wave, the photon's polarization signature is imprinted in the emission direction of the electron \cite{Kaaret2021}.  On the plane perpendicular to the photon propagation direction, the emission angle of the photoelectron has a $\cos^2$ distribution with respect to the polarization angle of the photon.  Thus, the X-ray polarization can be measured via electron tracking in a low density medium such as a gas. For PolarLight, this is done with a gas pixel detector (GPD), which is a 2D position-sensitive ionization chamber that enables the measurement of electron tracks. 

The GPD was first designed by the INFN-Pisa group and subsequently developed for space astronomy \cite{Costa2001,Bellazzini2006,Bellazzini2007b,Li2015}. It is a 2D gas proportional counter with pixel readout.  Dimethyl ether (DME; CH$_3$OCH$_3$) at a pressure of 0.8~atm was selected as the working gas, thanks to its low electron diffusion coefficient.  The gas has an absorption depth of 1~cm, and is sealed in a ceramic chamber with a 100~$\mu$m thick beryllium entrance window for incident X-rays.  After the DME absorbs an X-ray, a photoelectron is generated, with a kinetic energy slightly less than the photon energy (minus the electron binding energy, which is 543~eV for the O K-shell and less for C and H).  Then, the photoelectron travels in the gas and ionizes DME molecules, leaving behind pairs of electrons and ions along its trajectory.  A high voltage is applied in the chamber to create a drift electric field of 2~kV$/$cm, such that the ions move toward the cathode, which is the beryllium window, and the electrons move toward the anode.  The drift field strength is chosen to minimize the transversal diffusion of electrons \cite{Li2015}. 

\begin{figure}[t]
\centering
\includegraphics[width=0.7\columnwidth]{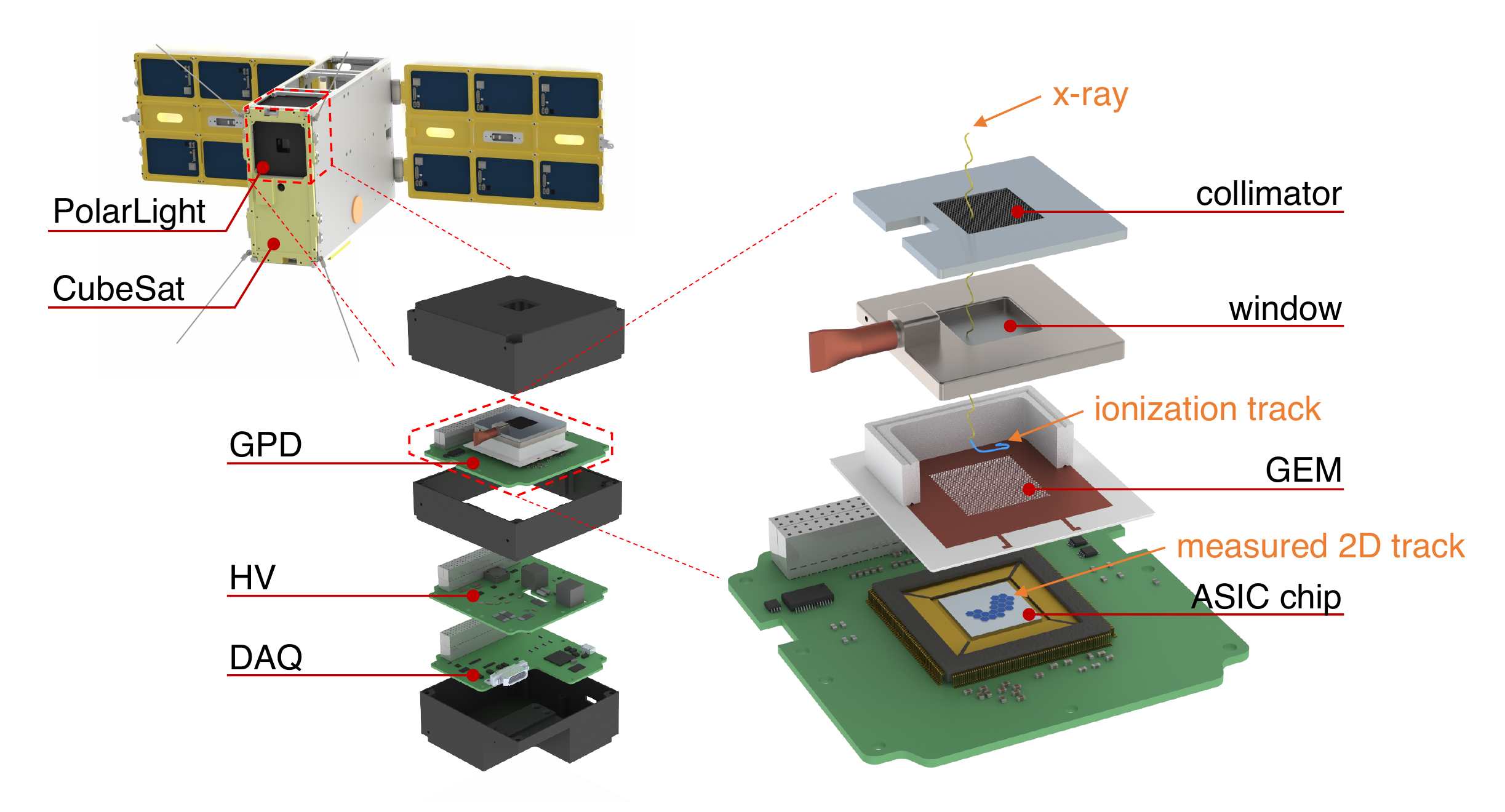} 
\includegraphics[width=0.29\columnwidth]{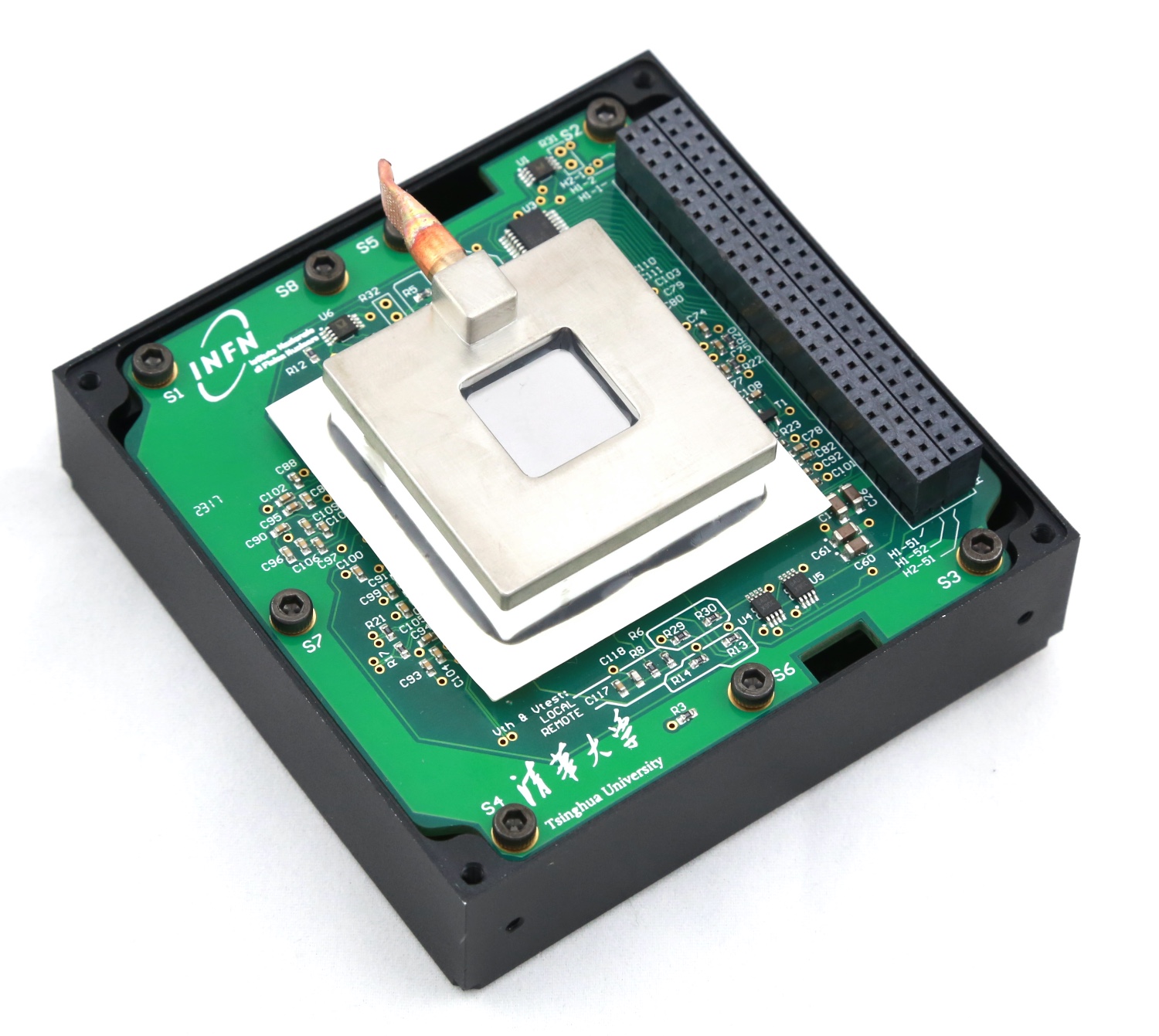}
\caption{Schematic drawing (left) and picture (right) of PolarLight.  PolarLight is a 1U payload in the CubeSat Tongchuan-1. It consists of three PCBs, hosting the GPD, HV, and DAQ, respectively, from top to bottom.  An exploded drawing of the GPD is shown to illustrate the detection of an event: an X-ray is absorbed after going through the collimator and window, a trajectory consisting of electrons and ions is created via ionizing the gas by the photoelectron, and a track image is detected when the electrons drift toward the ASIC.  Reproduced from Fig.~1 in Ref.~\cite{Feng2020}.}
\label{fig:le}
\end{figure}

In the chamber and right above the anode, there is a gas electron multiplier (GEM) for electron multiplication.  The GEM is a 100~$\mu$m thick  dielectric foil made of liquid crystal polymer (LCP) coated with 5 $\mu$m copper on both sides \cite{Tamagawa2009}, manufactured by SciEnergy Inc.  The foil consists of laser etched microholes with a diameter of 50 $\mu$m and a pitch of 100 $\mu$m in a hexagonal pattern.  A high voltage of about 600--700~V is applied across the top and bottom copper electrodes. The field lines are squeezed into the microholes, where the electric field strength is high enough to trigger avalanches when electrons go through the holes. 

Beneath the GEM lies an ASIC chip \cite{Bellazzini2006b}, which is the anode of the chamber used for collecting and processing the charge signals. The chip is pixelated to $300 \times 352$ pixels in a hexagonal pattern. Each pixel has a full electronic chain including the preamplifier, shaping amplifier, sample and hold, and multiplexer.  The noise is around 50 e$^-$ rms per pixel. The shaping time is 3--10 $\mu$s and externally adjustable.  The ASIC pixels measure the charges produced by the photoelectron after multiplication with the GEM, and provide the 2D photoelectron track on the plane of the detector. 

\subsection{Payload}

The PolarLight payload has a size of 1U, containing three printed circuit boards (PCBs) in an aluminum case. From top to bottom, the three PCBs respectively host the GPD, the high voltage (HV) power supply, and the data acquisition (DAQ) system. 

To avoid source confusion and reduce the diffuse background, a collimator is mounted on top of the GPD.  The collimator is a capillary plate made of lead glass (with $\sim$38\% lead oxide), manufactured by North Night Vision Technology Co.\ Ltd.  It is 1.66~mm thick and contains cylindrical micropores with a diameter of 83~$\mu$m and an open fraction of 71\%. The field of view (FOV) is $2.3^\circ$ full width at half maximum (FWHM), or $5.7^\circ$ full width at zero response.  

The HV is provided by a compact module UMHV0540N manufactured by HVM Technology, Inc. It has a cubic geometry with a length of 0.5 inch on each side, and is powered by a low voltage power supply of 5~V. With a programming pin, the HV output is adjustable from 0 to $-$4~kV. The output is split into three channels to power the drift plate and the two electrodes of the GEM.  The three HV outputs are monitored with analog to digital converters (ADCs). Two independent HV modules are mounted on the same board for a cold backup; the backup has never been used on orbit.

The DAQ board contains a microcontroller, ADC, flash memory, and communication interface to the CubeSat.  It also receives the global positioning system (GPS) signal for accurate timing. 

The total mass of PolarLight is about 580~g and the total power consumption is about 2.2~W during normal operation. 

\subsection{Performance}

\begin{figure}[t]
\centering
\includegraphics[width=0.49\columnwidth]{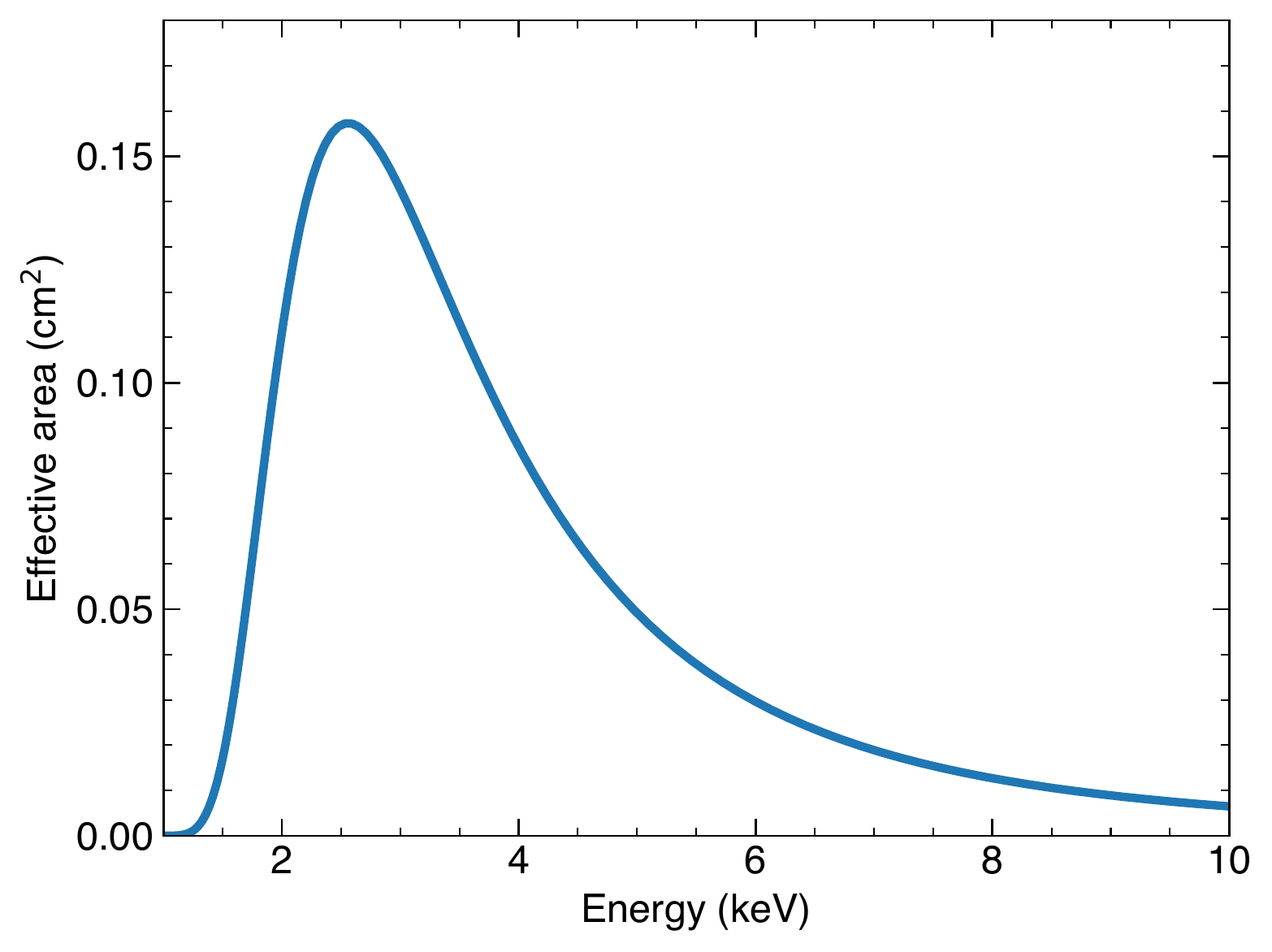}
\includegraphics[width=0.49\columnwidth]{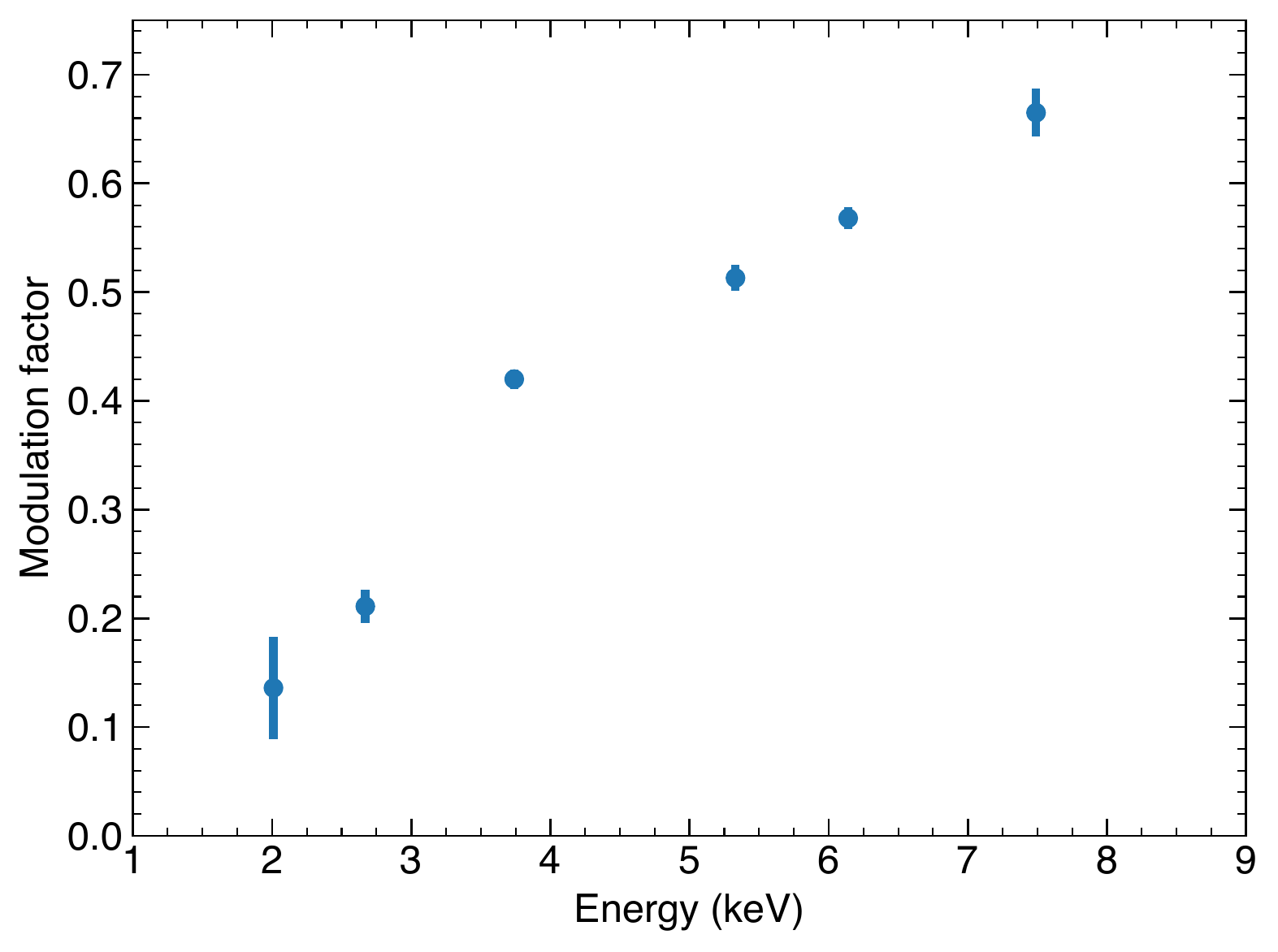}
\caption{Effective area (left) and modulation factor (right) of the PolarLight detector as a function of energy. Reproduced using data in Ref.~\cite{Feng2019}}
\label{fig:eff}
\end{figure}

The effective area of PolarLight as a function of energy is shown in Fig.~\ref{fig:eff}.  The high energy roll-over is determined by the quantum efficiency of the sensitive volume, i.e., 0.8~atm pure DME at a thickness of 1~cm.  The 100~$\mu$m beryllium window together with the thermal insulation foil outside the CubeSat shape the low-energy cutoff. Only X-rays absorbed in the region above the ASIC chip can be recorded; in practice, we discard events near the edge of the ASIC and only use the central $\pm7$~mm region for analysis. Therefore, the geometric detection area is 1.96~cm$^2$.  Considering the open fraction of the collimator, which is 71\%, and the detection efficiency, the effective area has a peak of about 0.16~cm$^2$. Therefore, PolarLight is indeed a tiny instrument. 

The modulation factor, which is the amplitude of the angular modulation of photoelectrons in response to fully polarized X-rays, determines the sensitivity of polarization measurement, and is one of the key parameters of the instrument.  The modulation factors at different energies in the PolarLight band, also displayed in Fig.~\ref{fig:eff}, are measured with 45$^\circ$ Bragg diffracted X-rays, which are 100\% polarized.  At low energies, the electron track is short and partially unresolved, leading to low modulation factors.

\begin{figure}[t]
\centering
\includegraphics[width=0.49\columnwidth]{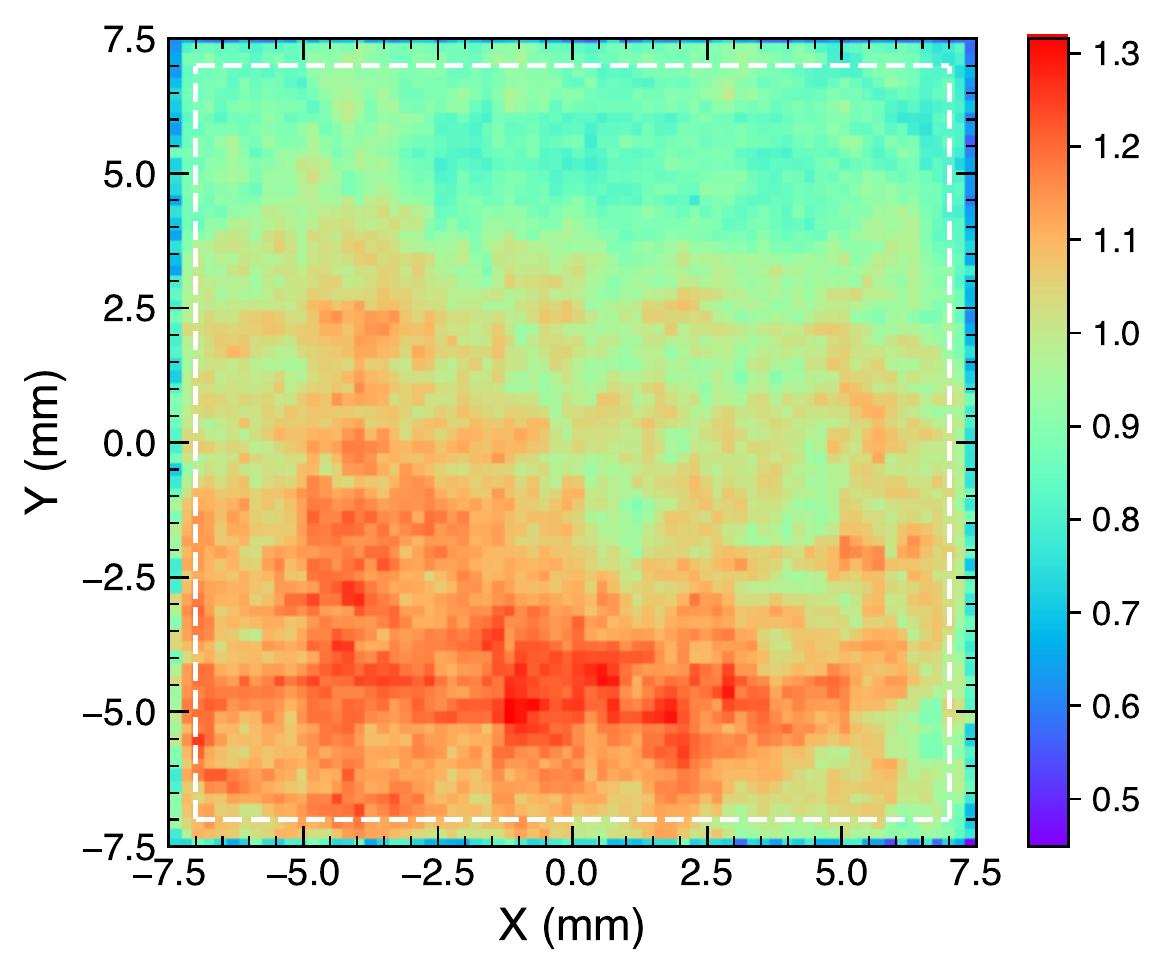}
\includegraphics[width=0.49\columnwidth]{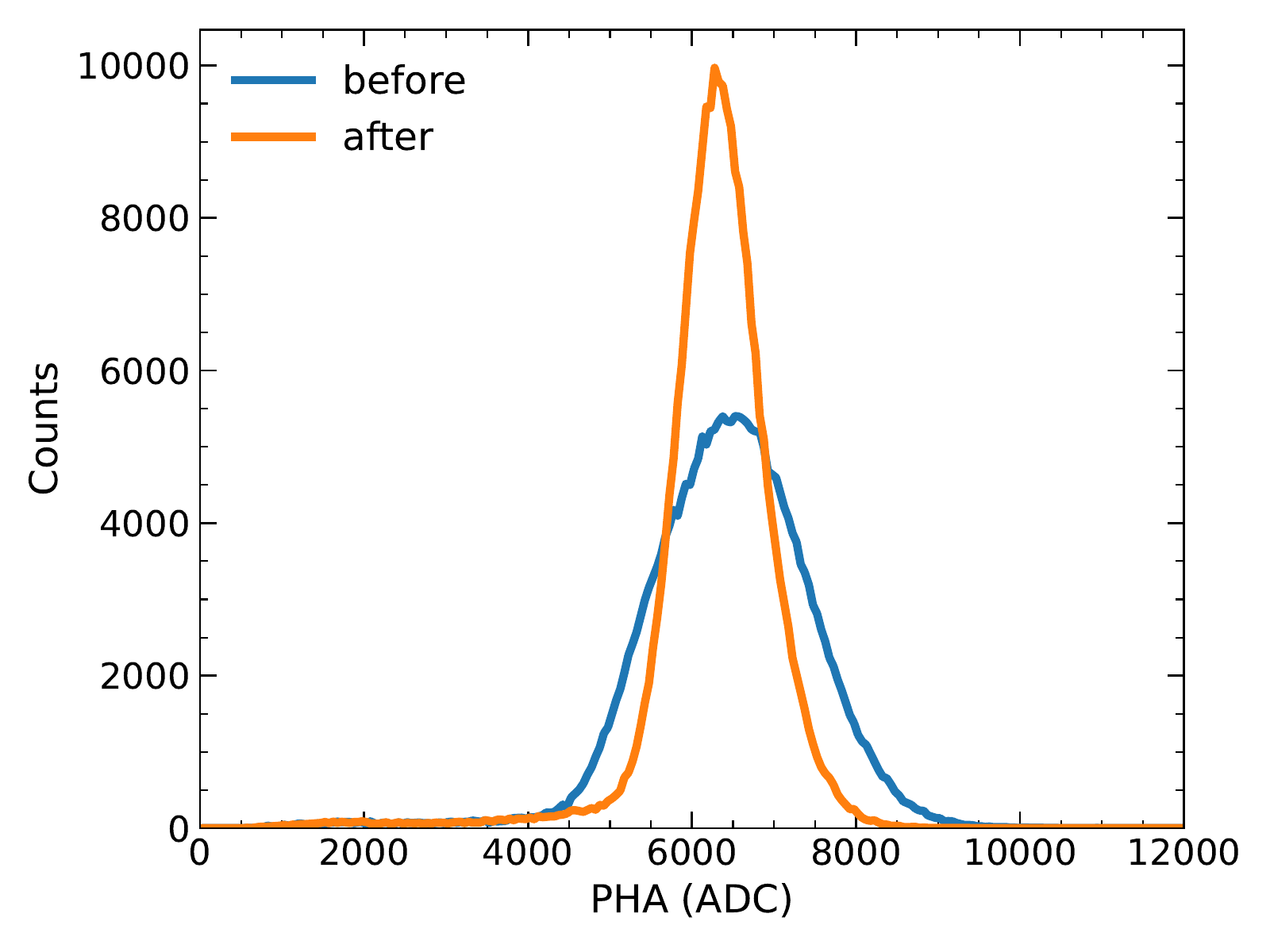}
\caption{Left: gain map measured with an $^{55}$Fe radioactive source.  The dashed lines encircle the central $\pm7$~mm region used for scientific analysis. Right: Energy spectra measured with $^{55}$Fe before (blue) and after gain map correction. Reproduced using data in Ref.~\cite{Feng2019}.}
\label{fig:gainmap}
\end{figure}

Due to the non-uniformity of the GEM thickness \cite{Takeuchi2014}, the GEM gain varies as a function of position and needs be calibrated and flattened. The gain map is measured with an $^{55}$Fe source illuminating the whole surface. The relative gain at different positions are measured and displayed in Fig.~\ref{fig:gainmap}. The spectral resolution (${\rm FWHM}/E$) of the detector at 5.9 keV improves from 33.5\% to 18.5\% after flat-fielding. 

\subsection{Operation}

\begin{figure}[t]
\centering
\includegraphics[width=0.9\columnwidth]{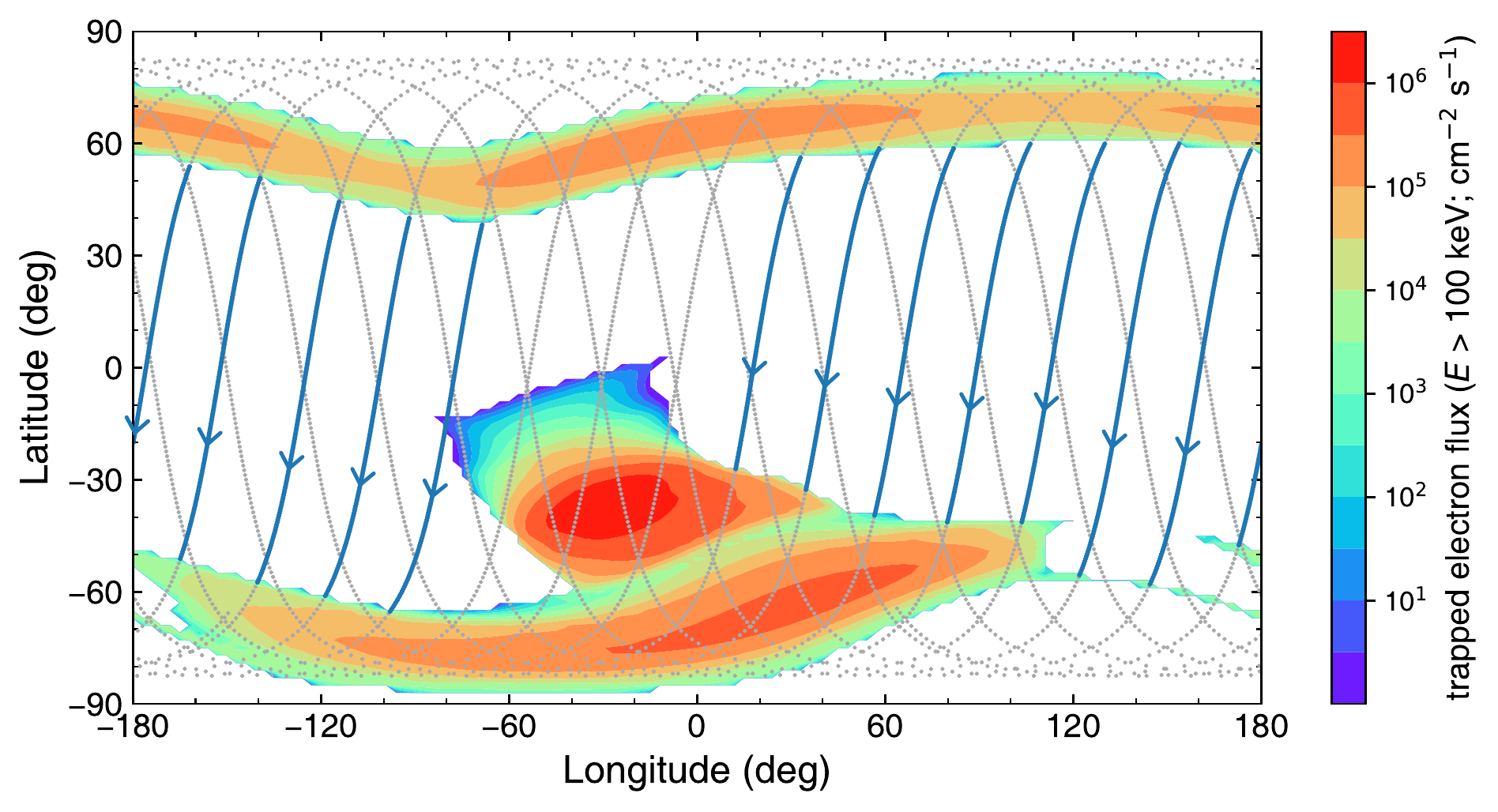}
\caption{A typical one-day orbit of PolarLight. The thick blue lines mark the time intervals when the target is visible.  The color map indicates the flux distribution of high energy ($>$100~keV) electrons trapped by the Earth's magnetosphere on the orbital plane, generated from SPENVIS (\url{http://www.spenvis.oma.be}). Reproduced using data in Ref.~\cite{Li2021}.}
\label{fig:orbit}
\end{figure}

PolarLight was launched into a LEO on October 29, 2018. It is a nearly circular Sun-synchronous orbit with an altitude of 520~km and a period of about 95~minutes.   Fig.~\ref{fig:orbit} shows a typical one-day orbit of the CubeSat on top of the flux map of high energy electrons trapped on the orbital plane by the Earth's magnetosphere.  As charged particles may damage the GPD if the flux is too high, the HV is powered off when the CubeSat passes through the high flux regions, including the south Atlantic anomaly (SAA) and two polar regions (Fig.~\ref{fig:orbit}).  As it takes time to power on/off the HV, we do not attempt to turn on the detector if the window for observation is less than 15 min, and discard the short observing windows near the two poles.  In addition, in around half of the durations suitable for observations, the target may be occulted by the Earth. On average, there are around 10 orbits a day, each with an effective exposure of 15 min or so, in which science data can be obtained.  

Limited by the small effective area, PolarLight is only sensitive to the brightest X-ray sources in the sky.  The Crab nebula is the primary target of PolarLight. It is also the only source with a significant detection in keV X-ray polarization prior to the launch of PolarLight \cite{Weisskopf1976,Weisskopf1978a}.  The second target is Scorpius (Sco) X-1, which is the first discovered extrasolar X-ray source and the brightest persistent object in this band besides the Sun. Transient X-ray binaries in the Milky Way with a brightness similar to the Crab nebula are also potential targets. 

\begin{figure}[t]
\centering
\includegraphics[width=0.9\columnwidth]{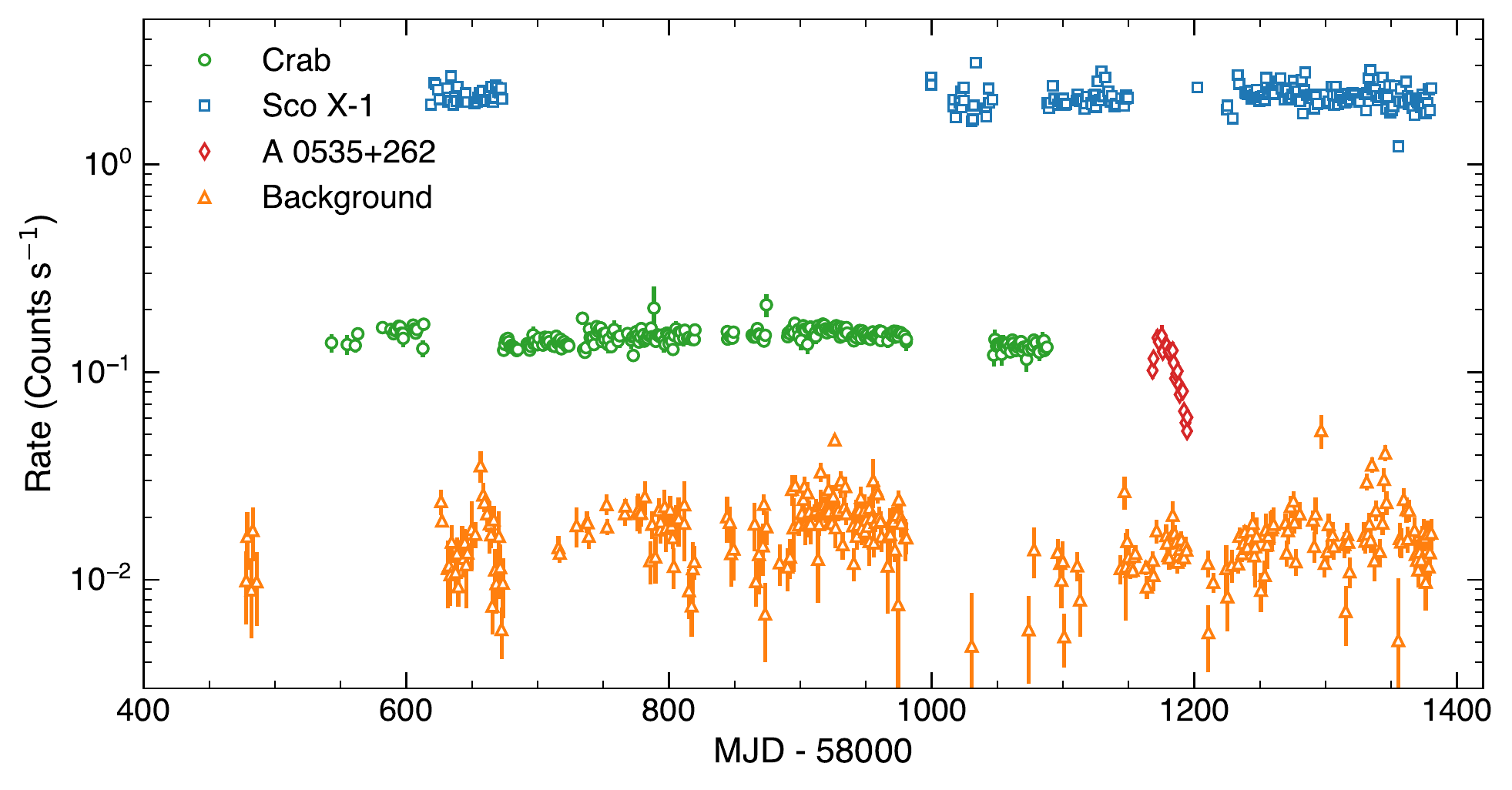}
\caption{One-day averaged lightcurves for the three targets and background observed with PolarLight. Reproduced using data in Ref.~\cite{Li2021}.}
\label{fig:lc}
\end{figure}

In the first four months after the launch,  the PolarLight detector was only briefly powered on for functional test, due to major software upgrades of the CubeSat and massive tests of the ground stations.  This long delay was caused by unexpected technical issues and could be significantly reduced.  Regular science observations started in March 2019 and the first target was the Crab nebula.  When the Crab nebula was not observable due to Sun avoidance, PolarLight observed Sco X-1.  In August 2020, the monitoring program for the Crab nebula ended and Sco X-1 became the primary target. In late 2020, the accreting pulsar A~0535+262 went into outburst, with a peak flux several times the Crab nebula. The schedule for Sco X-1 was interrupted in response to this target of opportunity.  In each orbit, during Earth occultation of the target, the instrument points at a random direction for background measurement. A one-day averaged light curve for the three sources as well as the background are shown in Fig.~\ref{fig:lc}.  The data archive of PolarLight as of October 2021 is summarized in Table~\ref{tab:obs}. 

\begin{table}
\caption{The data archive of PolarLight as of October 2021.}
\label{tab:obs}
\begin{tabular}{llll}
\hline\noalign{\smallskip}
Target\hspace{5em} & R.A. (J2000)\hspace{3em} & Decl. (J2000)\hspace{3em} & Exposure (ks) \\
\noalign{\smallskip}\hline\noalign{\smallskip}
Crab nebula & 05:34:31.9 & $+$22:00:52 & 1402  \\
Sco X-1 & 16:19:55.1 & $-$15:38:25 & 884 \\
A 0535+262 & 05:38:54.6 & $+$26:18:57 & 70 \\
Background & N.A. & N.A. & 393 \\
\noalign{\smallskip}\hline
\end{tabular}
\end{table}

\subsection{On-orbit Background}

PolarLight measures the on-orbit background while the science target is occulted by the Earth. In that case, the detector points roughly at the anti-Sun direction (so that the solar panel faces the Sun for battery charging) and rocks with a half angle of about 30$^\circ$. Therefore, a random sky region or the Earth's atmosphere is observed.  Taking into account the narrow FOV and small effective detection area, as well as the fact that the dominant source of background is charged particles on the orbit, pointing at a random direction is equivalent of observing the pure background.  

In order to model and understand the background, we created a mass model and ran particle simulations \cite{Huang2021} using the Geant4 package\footnote{\url{https://geant4.web.cern.ch}} . The mass model includes detailed structures of the GPD and the whole CubeSat.  The source of background consists of the diffuse cosmic X-rays \cite{Gruber1999} and high energy charged particles on the orbit \cite{Mizuno2004}.  

\begin{figure}[t]
\centering
\includegraphics[width=0.9\columnwidth]{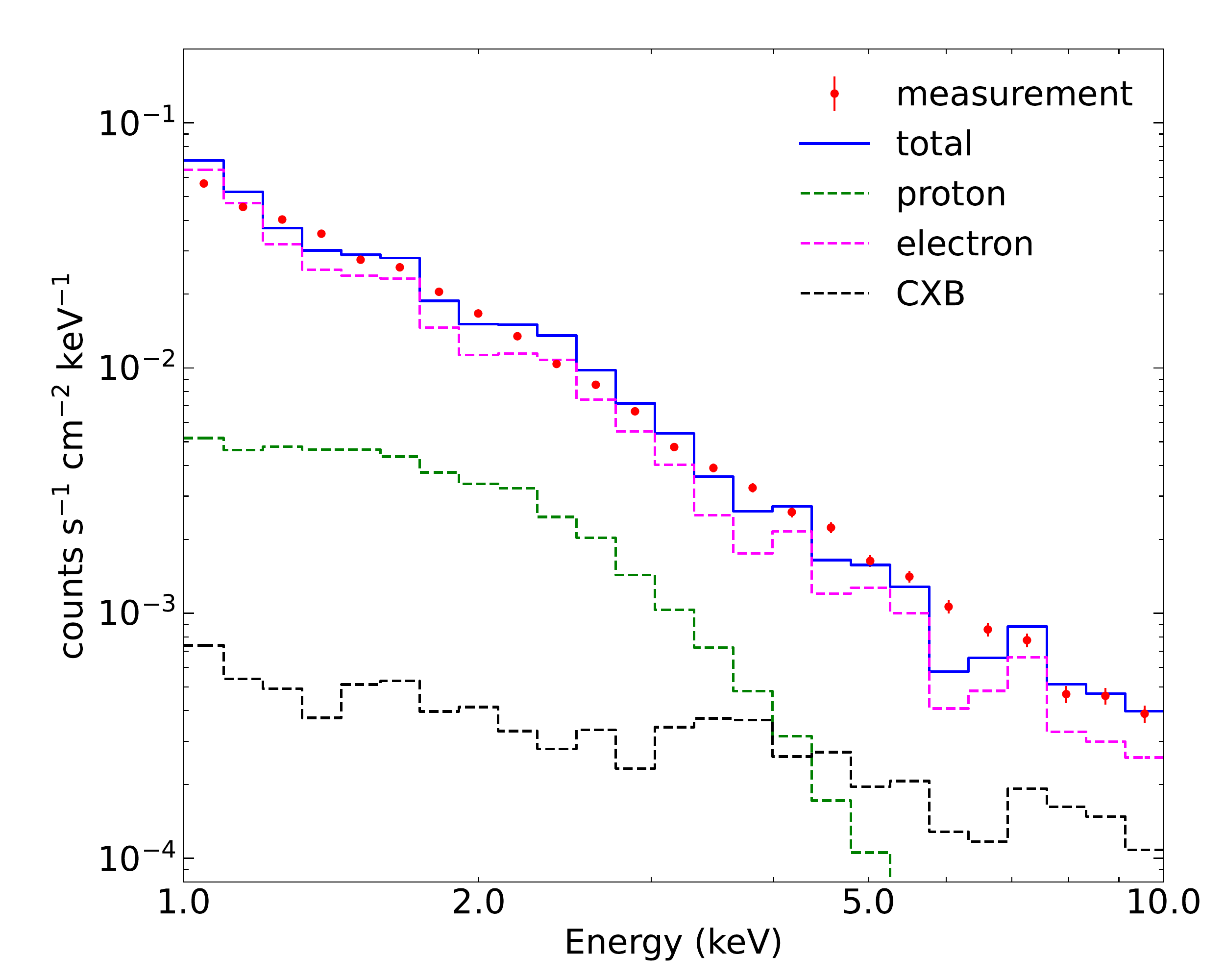}
\caption{Observed (dots) and simulated (lines) background spectra of PolarLight. Reproduced using data in Ref.~\cite{Huang2021}.}
\label{fig:bkg_spec}
\end{figure}

\begin{figure}[t]
\centering
\includegraphics[width=\columnwidth]{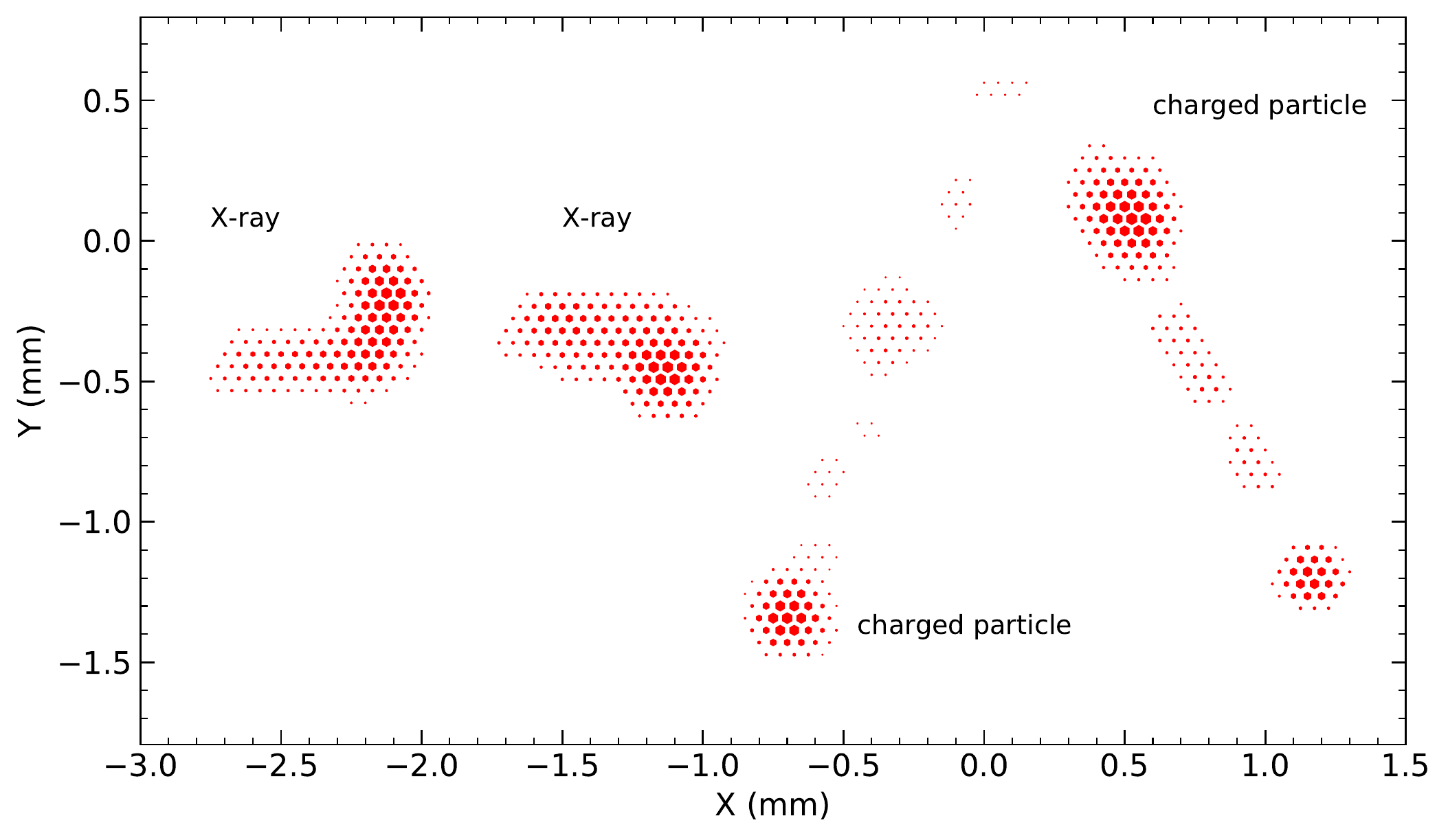}
\caption{Typical track images measured with PolarLight in the energy bin of 5.0--5.5~keV. Tracks due to X-ray events are usually small in size and show a curved morphology, while the charged particles tend to produce long and straight tracks. Reproduced using data in Ref.~\cite{Zhu2021}.}
\label{fig:track}
\end{figure}

The simulated background spectrum decomposed into different incident components is plotted in Fig.~\ref{fig:bkg_spec} against the measurement.  It is obvious that the electron-induced background is the dominant component. It occupies 81\% of the background counts in 1--10 keV, or 76\% in 2--8 keV. The fractions are 15\% and 17\% for the proton-induced background, and 4\% and 7\% for the leaked (i.e., not through the apertures) X-ray background, respectively, in the two energy bands.  

Events caused by X-rays of a few keV and those due to high energy charged particles may show different image morphologies. Generally speaking, the tracks produced by X-rays are shorter and more curved than those produced by high energy charged particles.  Event images that are likely resulted from X-rays and high energy charged particles, respectively, are displayed in Fig.~\ref{fig:track} for comparison.  This difference allows us to distinguish background events from source events. An energy dependent algorithm is proposed for particle discrimination \cite{Zhu2021}.  When an event triggers the electronics, a rectangle readout window is defined to include all triggered pixels plus some margins. The diagonal size of the readout window and the number of pixels above the triggering threshold, as a function of event energy, are used to distinguish between the source and background events, based on their different distributions.  They find that such a discrimination algorithm can remove roughly 70\% of the background events measured with PolarLight, while the remaining events show the same distributions as the source photons do, and cannot be removed.  This is not because the algorithm is not effective enough, rather it is due to the nature of energy deposited. Simulations~\cite{Huang2021} reveal that $\sim$30\% of the background events are due to energy deposits from secondary electrons with an energy of a few keV, i.e., these are essentially identical to the photoelectrons produced by X-rays and thus indistinguishable. Given the background level after discrimination, with an observing time of 1 Ms, PolarLight can reach a minimum detectable polarization of about 10\% for a 0.2 Crab source, or 5\% for a 0.5 Crab source. 

\subsection{Science results}

Prior to PolarLight, the Crab nebula was the only astrophysical source with a significant polarimetric measurement in the keV band.  In the 1970s, the Bragg polarimeter onboard OSO-8 measured a polarization fraction (PF) of $0.157 \pm 0.015$ with a polarization angle (PA) of $161.1^\circ \pm 2.8^\circ$ from the total nebular emission \cite{Weisskopf1976}; a phase-resolved analysis \cite{Weisskopf1978a} indicates that the pure nebular emission has a PF of $0.192 \pm 0.010$ and a PA of $156.4^\circ \pm 1.4^\circ$.  The above results were obtained in a narrow band around 2.6 keV due to the nature of Bragg diffraction, and consistent results were seen around 5.2 keV (the second order diffraction) with slightly larger uncertainties. 

With PolarLight, two new findings were obtained with observations of the Crab nebula.  In 2019, PolarLight revealed a possible variation in polarization (a sudden decrease in PF) coincident in time with the glitch of the Crab pulsar on July 23 \cite{Feng2020a}. The variation was found to have a significance of 3$\sigma$ using different methods, including the Bayes factors, Bayesian posterior distributions, and bootstrap analysis. This may suggest that the pulsar magnetosphere altered after the glitch.  Then,  the polarization recovered roughly 100 days after the glitch \cite{Long2021}. With more data being accumulated, the PA measured with PolarLight from the total nebular emission was found to have a difference of $18.0^\circ \pm 4.6^\circ$ from that measured 42 years ago with OSO-8, indicating a secular evolution of polarization associated with either the Crab nebula or pulsar \cite{Long2021}.  The long-term variation in PA could be a result of multiple glitches in the history \citep{Feng2020a,Espinoza2011},  magnetic reconnection in the synchrotron emitting regions in the nebula \cite{Moran2013,Moran2016}, or secular evolution of the pulsar magnetic geometry \cite{Lyne2013,Ge2016}. 

Sco X-1 is the brightest persistent extrasolar object in the keV sky. It is powered by accretion from a low-mass companion star onto a low-magnetic neutron star \cite{Steeghs2002}, classified to be a so-called Z-source based on the color-color diagram \cite{Hasinger1989}.  OSO-8 observed the source and produced a non-detection in X-ray polarization around 2.6~keV, but obtained a 3$\sigma$ detection around 5.2~keV \cite{Long1979}.  PolarLight yielded consistent results: a non-detection below 4 keV but a significant detection in 4--8 keV; the significance in the hard band is up to 5$\sigma$ when the source shows the highest intensity \cite{Long2022}. The PA measured with PolarLight is consistent with that obtained with OSO-8 within errors, and in line with the orientation of the radio jet of Sco X-1 on the plane of the sky \cite{Fomalont2001}.  The jet is supposed to be launched along the symmetry axis of the system. The results favor the scenario that an optically-thin corona is located in the transition layer of Sco X-1 under the highest accretion rates \cite{Revnivtsev2006,Dai2007,Titarchuk2014}, and disfavor the accretion disk corona model \cite{Barnard2003,Church2012}. 

A 0535+262 is an accreting pulsar and underwent a giant outburst in 2020, with a peak flux of about 3~Crab in the PolarLight band \cite{Kong2021}. PolarLight monitored the outburst and produced a non-detection in X-ray polarization with only an upper limit. The results are being analyzed and will be reported elsewhere in the near future (Long et al.\ {\it in prep.}). 

\begin{figure}[t]
\centering
\includegraphics[width=0.49\columnwidth]{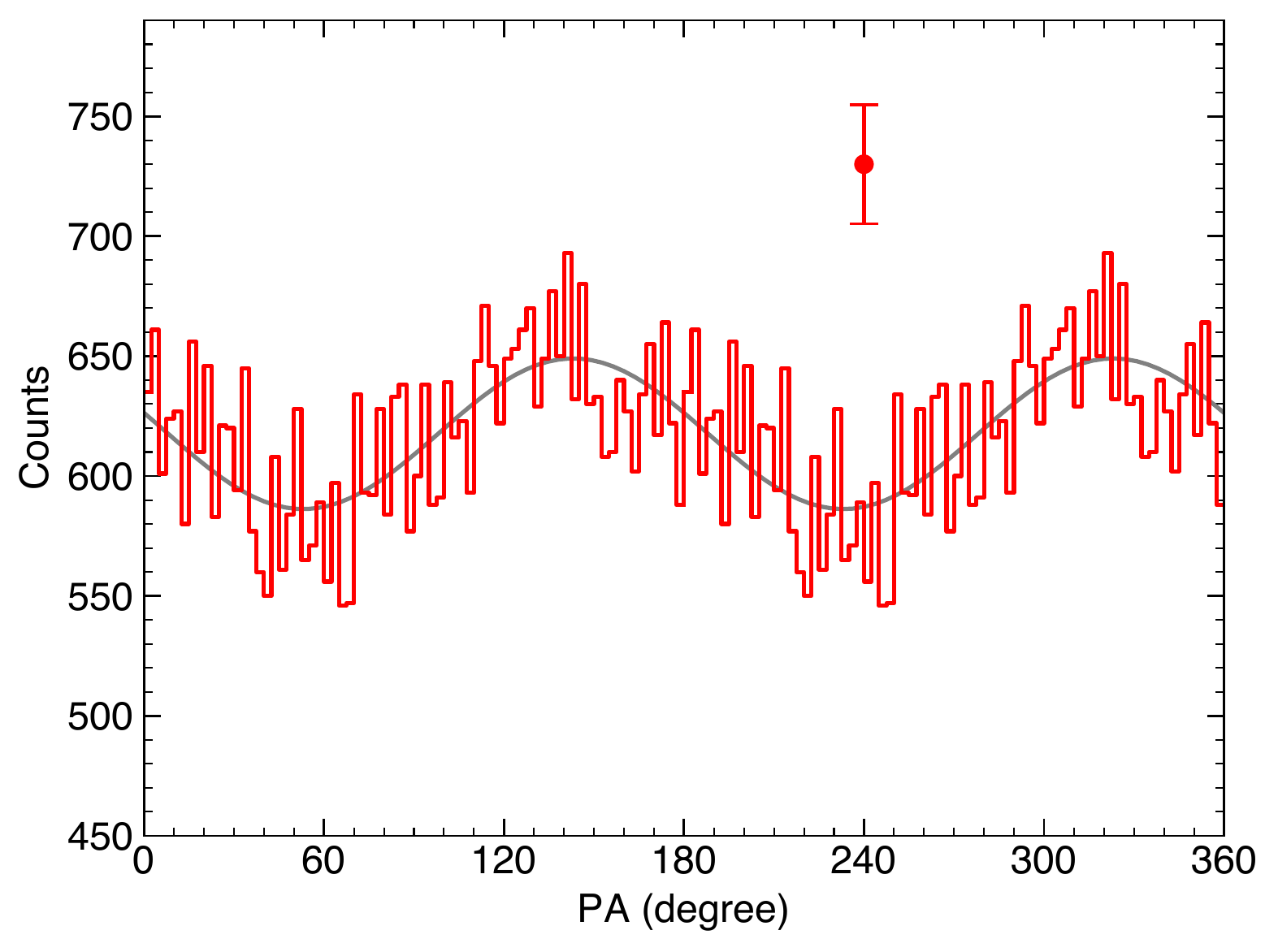}
\includegraphics[width=0.49\columnwidth]{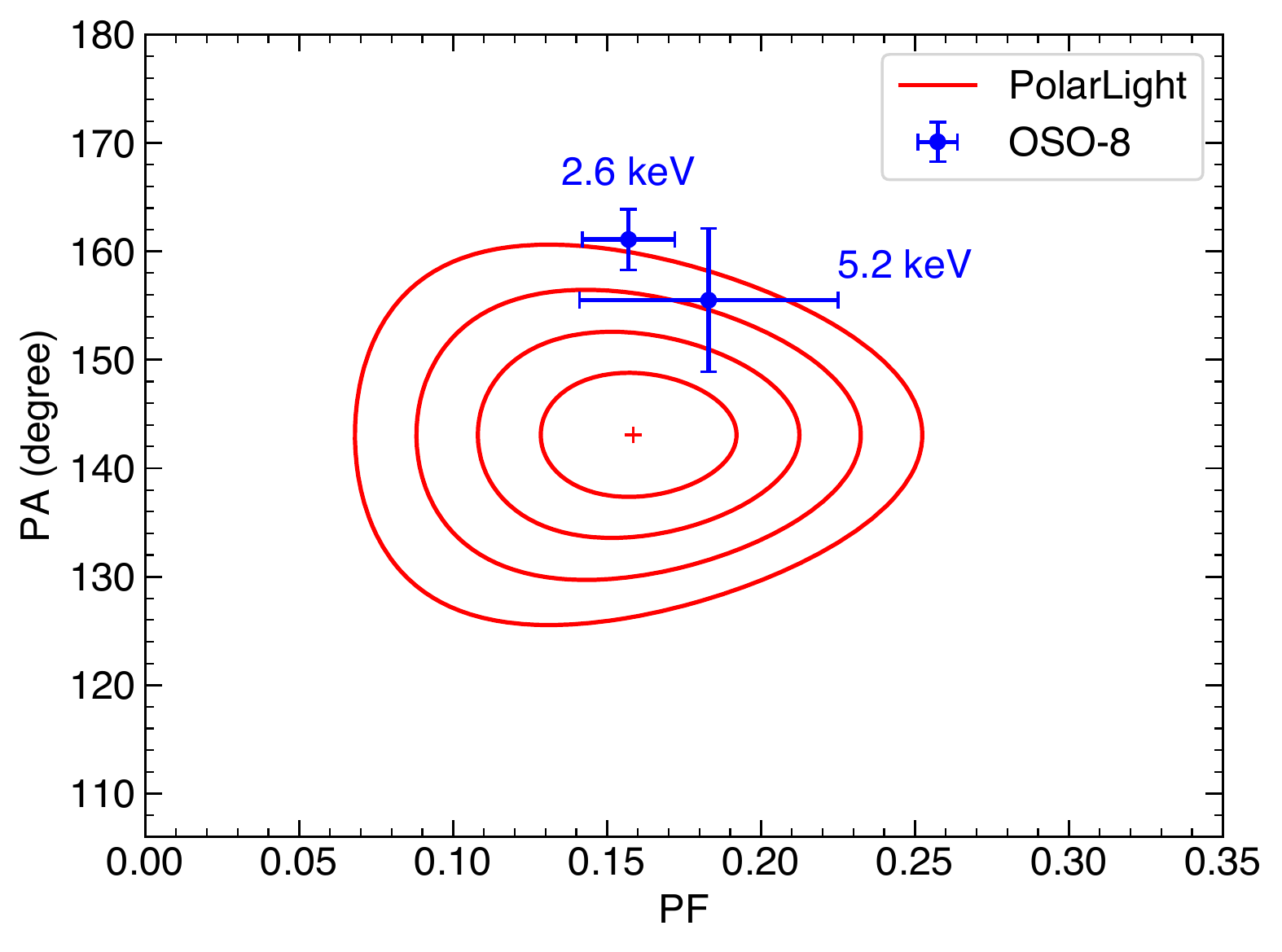}
\caption{Left: polarization modulation curve of the Crab nebula measured with PolarLight in the energy range of 3--4.5 keV. The error bar indicates the typical error and the gray curve represents the model. Right: PA vs.\ PF posterior contours derived from Bayesian analysis. The blue points are the OSO-8 measurements around 2.6~keV and 5.2~keV, respectively. Reproduced using data in Refs.~\cite{Long2021}. }
\label{fig:crab}
\end{figure}

\section{Discussion}
\label{sec:dis}

HaloSat and PolarLight demonstrate that CubeSats can be useful platforms for X-ray astronomy. CubeSats can produce cutting-edge science results.  They provide affordable access to space for scientific research which is particularly important for universities and colleges, where students may take a leading role.

CubeSats are not just scaled-down versions of large satellites. Due to the high risk of space exploration and the huge investment behind large space missions, reliability always has first priority.  Therefore, strict rules are required in the choice of components and materials, and extensive documentation, testing, and review is required in each development phase. These incur considerable cost in both time and money. For CubeSat-based missions, one may use strategies to decrease cost and schedule while incurring higher, but tolerable, risk. 

One such strategy is the use of commercial parts. Most CubeSats are launched in to LEO which presents a relatively benign radiation environment and have mission durations of a few years or less, limiting the radiation dose. This enables use of commercial parts, including those that have not been tested in space. Commercial parts can offer significant cost saving and decreased procurement times. Selection of an automotive-grade microcontroller for the PolarLight DAQ board provides an example of an appropriate balance between cost and reliability. This strategy can be extended to the use of larger components such as single board computers (SBCs), amplifiers, and power supplies. This also saves significant development time and cost. It is important to perform environmental testing before final selection for flight of such parts, including extensive thermal cycling, thermal-vacuum testing, and radiation testing. A commercial high voltage power supply (HVPS) was selected for PolarLight, based on its compactness. The HVPS was then subjected to thermal and thermal-vacuum tests. Also, a cold back-up system was implemented in case of failure.


Reduction in required documentation and reviews can greatly streamline the development process. Typically, only the key requirements and interfaces need be documented. CubeSat development can bypass the initial study phases of larger missions which typically produce mainly documentation. The time saved can be put into building and testing of prototypes, which also puts more joy in the process. Reviews by experts outside the project team should be performed, but can be advisory rather than act as gates to program continuation. This makes the reviews more collaborative, with the review team invested in and actively contributing to mission success.

Software development (flight, operations, and data processing) for CubeSats is challenging because the mission teams are quite small while the complexity of the needed software is not necessarily reduced by the same factor. Re-use or adaption of existing software is essential. HaloSat used an existing system for mission operations developed by Blue Canyon Technologies and new software was written only for HaloSat-specific observation planning. HaloSat adapted previously developed codes to calculate the silicon drift detector response and PolarLight used previously developed codes to analyze photoelectron tracks. Flight software should be made as simple as possible, but not simpler (to paraphrase Einstein) and extensively tested. For both HaloSat and PolarLight, the payload computer is rebooted each spacecraft orbit to bypass bugs (such as memory leaks) that accumulate in long-term running and to limit the impact of software crashes to at most a single orbit. One advantage of CubeSats regarding data processing software is that the data volumes are typically limited. Therefore, processing efficiency is not paramount allowing use of higher level languages such as Python. 


We note that PolarLight is different from HaloSat and many other missions, in the sense that it is not the sole payload and shares the CubeSat with others. This means that we just need to pay for a ``seat'' instead of a whole ``bus'', which lowers the cost substantially. Of course, it is important to ensure there is no conflict between payloads, including regarding mission operations.  PolarLight is the only payload on its spacecraft that requires specific pointing and, in other words, was able to obtain the ``driver's seat''.  However, the spacecraft on-board computer (OBC) did not allow for customized programs to maximize the observation efficiency. Observations with PolarLight were accomplished via combinations of commands uploaded daily.

CubeSats are a great tool for student training. The development of large-scale missions usually takes a decade or so.  It is unlikely for a student to follow the project from the very beginning to the end.  It is also hard for students to take leading roles in large projects.  In contrast, both PolarLight and HaloSat were developed in less than two years and with involvement of individual students from near project initiation to on-orbit operations. Students played lead roles in the construction and testing of engineering and flight model instruments and in the flight-model scientific calibration. Students also played lead roles in the flight operations and the development of mission planning and data processing software. The time scale is sufficiently short that the student on HaloSat who acquired the first X-ray spectra with the engineering model detectors was able to write a thesis using data collected during the first year of on-orbit operations. Students can mature quickly if they are provided with a major role on a real space mission. Such training is a rare opportunity but can be well adapted into a CubeSat project.  Such training extends to science, engineering, and leadership.


%


\begin{thebibliography}{10}

\bibitem{Shkolnik2018}
E.~L. {Shkolnik}, ``{On the verge of an astronomy CubeSat revolution},'' {\em
  Nature Astronomy}, vol.~2, pp.~374--378, May 2018.

\bibitem{Kaaret2019}
P.~{Kaaret}, A.~{Zajczyk}, D.~M. {LaRocca}, {\em et~al.}, ``{HaloSat: A CubeSat
  to Study the Hot Galactic Halo},'' {\em \apj}, vol.~884, p.~162, Oct. 2019.

\bibitem{Feng2019}
H.~{Feng}, W.~{Jiang}, M.~{Minuti}, {\em et~al.}, ``{PolarLight: a CubeSat
  X-ray polarimeter based on the gas pixel detector},'' {\em Experimental
  Astronomy}, vol.~47, pp.~225--243, Apr. 2019.

\bibitem{Moore2018}
C.~S. {Moore}, A.~{Caspi}, T.~N. {Woods}, {\em et~al.}, ``{The Instruments and
  Capabilities of the Miniature X-Ray Solar Spectrometer (MinXSS) CubeSats},''
  {\em \solphys}, vol.~293, p.~21, Feb. 2018.

\bibitem{Mason2020}
J.~P. {Mason}, T.~N. {Woods}, P.~C. {Chamberlin}, {\em et~al.}, ``{MinXSS-2
  CubeSat mission overview: Improvements from the successful MinXSS-1
  mission},'' {\em Advances in Space Research}, vol.~66, pp.~3--9, July 2020.

\bibitem{Kaaret2020}
P.~{Kaaret}, D.~{Koutroumpa}, K.~D. {Kuntz}, {\em et~al.}, ``{A disk-dominated
  and clumpy circumgalactic medium of the Milky Way seen in X-ray emission},''
  {\em Nature Astronomy}, vol.~4, pp.~1072--1077, Oct. 2020.

\bibitem{Spitzer1956}
J.~{Spitzer}, Lyman, ``{On a Possible Interstellar Galactic Corona.},'' {\em
  \apj}, vol.~124, p.~20, July 1956.

\bibitem{Putman2012}
M.~E. {Putman}, J.~E.~G. {Peek}, and M.~R. {Joung}, ``{Gaseous Galaxy Halos},''
  {\em \araa}, vol.~50, pp.~491--529, Sept. 2012.

\bibitem{Tumlinson2017}
J.~{Tumlinson}, M.~S. {Peeples}, and J.~K. {Werk}, ``{The Circumgalactic
  Medium},'' {\em \araa}, vol.~55, pp.~389--432, Aug. 2017.

\bibitem{Shull2012}
J.~M. {Shull}, B.~D. {Smith}, and C.~W. {Danforth}, ``{The Baryon Census in a
  Multiphase Intergalactic Medium: 30\% of the Baryons May Still be Missing},''
  {\em \apj}, vol.~759, p.~23, Nov. 2012.

\bibitem{BlandHawthorn2016}
J.~{Bland-Hawthorn} and O.~{Gerhard}, ``{The Galaxy in Context: Structural,
  Kinematic, and Integrated Properties},'' {\em \araa}, vol.~54, pp.~529--596,
  Sept. 2016.

\bibitem{Kuntz2019}
K.~D. {Kuntz}, ``{Solar wind charge exchange: an astrophysical nuisance},''
  {\em Astronomy and Astrophysics Review}, vol.~27, p.~1, Jan. 2019.

\bibitem{Kaaret2018}
P.~{Kaaret}, ``{Asking a big question with a small satellite},'' {\em Nature
  Astronomy}, vol.~2, pp.~755--755, Sept. 2018.

\bibitem{LaRocca2020}
D.~M. {LaRocca}, P.~{Kaaret}, K.~D. {Kuntz}, {\em et~al.}, ``{An Analysis of
  the North Polar Spur Using HaloSat},'' {\em \apj}, vol.~904, p.~54, Nov.
  2020.

\bibitem{Zajczyk2020}
A.~{Zajczyk}, P.~{Kaaret}, D.~{LaRocca}, {\em et~al.}, ``{On-ground calibration
  of the HaloSat science instrument},'' {\em Journal of Astronomical
  Telescopes, Instruments, and Systems}, vol.~6, p.~044005, Oct. 2020.

\bibitem{Scholze2009}
F.~{Scholze} and M.~{Procop}, ``{Modelling the response function of energy
  dispersive X-ray spectrometers with silicon detectors},'' {\em X-ray
  Spectrometry}, vol.~38, pp.~312--321, July 2009.

\bibitem{Gendreau2016}
K.~C. {Gendreau}, Z.~{Arzoumanian}, P.~W. {Adkins}, {\em et~al.}, ``{The
  Neutron star Interior Composition Explorer (NICER): design and
  development},'' in {\em Space Telescopes and Instrumentation 2016:
  Ultraviolet to Gamma Ray} (J.-W.~A. {den Herder}, T.~{Takahashi}, and
  M.~{Bautz}, eds.), vol.~9905 of {\em Society of Photo-Optical Instrumentation
  Engineers (SPIE) Conference Series}, p.~99051H, July 2016.

\bibitem{Arnaud1996}
K.~A. {Arnaud}, ``{XSPEC: The First Ten Years},'' in {\em Astronomical Data
  Analysis Software and Systems V} (G.~H. {Jacoby} and J.~{Barnes}, eds.),
  vol.~101 of {\em Astronomical Society of the Pacific Conference Series},
  p.~17, Jan. 1996.

\bibitem{Bluem2020}
J.~{Bluem}, P.~{Kaaret}, W.~{Fuelberth}, {\em et~al.}, ``{A HaloSat Analysis of
  the Cygnus Superbubble},'' {\em \apj}, vol.~905, p.~91, Dec. 2020.

\bibitem{Silich2020}
E.~M. {Silich}, P.~{Kaaret}, A.~{Zajczyk}, {\em et~al.}, ``{Global X-Ray
  Properties of the Vela and Puppis A Supernova Remnants},'' {\em \aj},
  vol.~160, p.~20, July 2020.

\bibitem{Ringuette2021}
R.~{Ringuette}, D.~{Koutroumpa}, K.~D. {Kuntz}, {\em et~al.}, ``{HaloSat
  Observations of Heliospheric Solar Wind Charge Exchange},'' {\em \apj},
  vol.~918, p.~41, Sept. 2021.

\bibitem{Gulick2021}
H.~{Gulick}, P.~{Kaaret}, A.~{Zajczyk}, {\em et~al.}, ``{Total X-Ray Emission
  from the LMC Observed with HaloSat},'' {\em \aj}, vol.~161, p.~57, Feb. 2021.

\bibitem{Hewitt2021}
N.~H. {Hewitt}, P.~{Kaaret}, and C.~A. {Fuller}, ``{HaloSat Observation of the
  Virgo Intracluster Medium},'' {\em Research Notes of the American
  Astronomical Society}, vol.~5, p.~185, Aug. 2021.

\bibitem{Silich2021}
E.~M. {Silich}, K.~{Jahoda}, L.~{Angelini}, {\em et~al.}, ``{A Search for the
  3.5 keV Line from the Milky Way's Dark Matter Halo with HaloSat},'' {\em
  \apj}, vol.~916, p.~2, July 2021.

\bibitem{Kaaret2021}
P.~{Kaaret}, ``{X-ray Polarimetry},'' in
  {\em The WSPC Handbook of Astronomical Instrumentation} (David N.\ Burrows, 
  ed.), vol.~4, pp.~281--300, July 2021 (arXiv:1408.5899). 

\bibitem{Costa2001}
E.~{Costa}, P.~{Soffitta}, R.~{Bellazzini}, {\em et~al.}, ``{An efficient
  photoelectric X-ray polarimeter for the study of black holes and neutron
  stars},'' {\em \nat}, vol.~411, pp.~662--665, June 2001.

\bibitem{Bellazzini2006}
R.~{Bellazzini}, F.~{Angelini}, L.~{Baldini}, {\em et~al.}, ``{Gas pixel
  detectors for X-ray polarimetry applications},'' {\em Nuclear Instruments and
  Methods in Physics Research A}, vol.~560, pp.~425--434, May 2006.

\bibitem{Bellazzini2007b}
R.~{Bellazzini}, G.~{Spandre}, M.~{Minuti}, {\em et~al.}, ``{A sealed Gas Pixel
  Detector for X-ray astronomy},'' {\em Nuclear Instruments and Methods in
  Physics Research A}, vol.~579, pp.~853--858, Sept. 2007.

\bibitem{Li2015}
H.~{Li}, H.~{Feng}, F.~{Muleri}, {\em et~al.}, ``{Assembly and test of the gas
  pixel detector for X-ray polarimetry},'' {\em Nuclear Instruments and Methods
  in Physics Research A}, vol.~804, pp.~155--162, Dec. 2015.

\bibitem{Feng2020}
H.~{Feng} and R.~{Bellazzini}, ``{The X-ray polarimetry window reopens},'' {\em
  Nature Astronomy}, vol.~4, pp.~547--547, May 2020.

\bibitem{Tamagawa2009}
T.~{Tamagawa}, A.~{Hayato}, F.~{Asami}, {\em et~al.}, ``{Development of
  thick-foil and fine-pitch GEMs with a laser etching technique},'' {\em
  Nuclear Instruments and Methods in Physics Research A}, vol.~608,
  pp.~390--396, Sept. 2009.

\bibitem{Bellazzini2006b}
R.~{Bellazzini}, G.~{Spandre}, M.~{Minuti}, {\em et~al.}, ``{Direct reading of
  charge multipliers with a self-triggering CMOS analog chip with 105 k pixels
  at 50 {\ensuremath{\mu}}m pitch},'' {\em Nuclear Instruments and Methods in
  Physics Research A}, vol.~566, pp.~552--562, Oct. 2006.

\bibitem{Takeuchi2014}
Y.~{Takeuchi}, T.~{Kitaguchi}, A.~{Hayato}, {\em et~al.}, ``{Properties of the
  flight model gas electron multiplier for the GEMS mission},'' in {\em Space
  Telescopes and Instrumentation 2014: Ultraviolet to Gamma Ray}
  (T.~{Takahashi}, J.-W.~A. {den Herder}, and M.~{Bautz}, eds.), vol.~9144 of
  {\em Society of Photo-Optical Instrumentation Engineers (SPIE) Conference
  Series}, p.~91444N, July 2014.

\bibitem{Li2021}
H.~{Li}, X.~{Long}, H.~{Feng}, {\em et~al.}, ``{In-orbit operation and
  performance of the CubeSat Soft X-ray polarimeter PolarLight},'' {\em
  Advances in Space Research}, vol.~67, pp.~708--714, Jan. 2021.

\bibitem{Weisskopf1976}
M.~C. {Weisskopf}, G.~G. {Cohen}, H.~L. {Kestenbaum}, {\em et~al.},
  ``{Measurement of the X-ray polarization of the Crab nebula.},'' {\em \apjl},
  vol.~208, pp.~L125--L128, Sept. 1976.

\bibitem{Weisskopf1978a}
M.~C. {Weisskopf}, E.~H. {Silver}, H.~L. {Kestenbaum}, K.~S. {Long}, and
  R.~{Novick}, ``{A precision measurement of the X-ray polarization of the Crab
  Nebula without pulsar contamination.},'' {\em \apjl}, vol.~220,
  pp.~L117--L121, Mar. 1978.

\bibitem{Huang2021}
J.~{Huang}, H.~{Feng}, H.~{Li}, {\em et~al.}, ``{Modeling the in-orbit
  Background of PolarLight},'' {\em \apj}, vol.~909, p.~104, Mar. 2021.

\bibitem{Gruber1999}
D.~E. {Gruber}, J.~L. {Matteson}, L.~E. {Peterson}, and G.~V. {Jung}, ``{The
  Spectrum of Diffuse Cosmic Hard X-Rays Measured with HEAO 1},'' {\em \apj},
  vol.~520, pp.~124--129, July 1999.

\bibitem{Mizuno2004}
T.~{Mizuno}, T.~{Kamae}, G.~{Godfrey}, {\em et~al.}, ``{Cosmic-Ray Background
  Flux Model Based on a Gamma-Ray Large Area Space Telescope Balloon Flight
  Engineering Model},'' {\em \apj}, vol.~614, pp.~1113--1123, Oct. 2004.

\bibitem{Zhu2021}
J.~{Zhu}, H.~{Li}, H.~{Feng}, {\em et~al.}, ``{Discrimination of background
  events in the PolarLight X-ray polarimeter},'' {\em Research in Astronomy and 
  Astrophysics}, vol.~21, p.~233, Nov. 2021.

\bibitem{Feng2020a}
H.~{Feng}, H.~{Li}, X.~{Long}, {\em et~al.}, ``{Re-detection and a possible
  time variation of soft X-ray polarization from the Crab},'' {\em Nature
  Astronomy}, vol.~4, pp.~511--516, May 2020.

\bibitem{Long2021}
X.~{Long}, H.~{Feng}, H.~{Li}, {\em et~al.}, ``{X-Ray Polarimetry of the Crab
  Nebula with PolarLight: Polarization Recovery after the Glitch and a Secular
  Position Angle Variation},'' {\em \apjl}, vol.~912, p.~L28, May 2021.

\bibitem{Espinoza2011}
C.~M. {Espinoza}, A.~G. {Lyne}, B.~W. {Stappers}, and M.~{Kramer}, ``{A study
  of 315 glitches in the rotation of 102 pulsars},'' {\em \mnras}, vol.~414,
  pp.~1679--1704, June 2011.

\bibitem{Moran2013}
P.~{Moran}, A.~{Shearer}, R.~P. {Mignani}, {\em et~al.}, ``{Optical polarimetry
  of the inner Crab nebula and pulsar},'' {\em \mnras}, vol.~433,
  pp.~2564--2575, Aug. 2013.

\bibitem{Moran2016}
P.~{Moran}, G.~{Kyne}, C.~{Gouiff{\`e}s}, {\em et~al.}, ``{A recent change in
  the optical and {\ensuremath{\gamma}}-ray polarization of the Crab nebula and
  pulsar},'' {\em \mnras}, vol.~456, pp.~2974--2981, Mar. 2016.

\bibitem{Lyne2013}
A.~{Lyne}, F.~{Graham-Smith}, P.~{Weltevrede}, {\em et~al.}, ``{Evolution of
  the Magnetic Field Structure of the Crab Pulsar},'' {\em Science}, vol.~342,
  pp.~598--601, Nov. 2013.

\bibitem{Ge2016}
M.~Y. {Ge}, L.~L. {Yan}, F.~J. {Lu}, {\em et~al.}, ``{Evolution of the X-Ray
  Profile of the Crab Pulsar},'' {\em \apj}, vol.~818, p.~48, Feb. 2016.

\bibitem{Steeghs2002}
D.~{Steeghs} and J.~{Casares}, ``{The Mass Donor of Scorpius X-1 Revealed},''
  {\em \apj}, vol.~568, pp.~273--278, Mar. 2002.

\bibitem{Hasinger1989}
G.~{Hasinger} and M.~{van der Klis}, ``{Two patterns of correlated X-ray timing
  and spectral behaviour in low-mass X-ray binaries.},'' {\em \aap}, vol.~225,
  pp.~79--96, Nov. 1989.

\bibitem{Long1979}
K.~S. {Long}, G.~A. {Chanan}, W.~H.~M. {Ku}, and R.~{Novick}, ``{The linear
  X-ray polarization of Scorpius X-1.},'' {\em \apjl}, vol.~232,
  pp.~L107--L110, Sept. 1979.

\bibitem{Long2022}
X.~{Long}, H.~{Feng}, H.~{Li}, {\em et~al.}, ``{A significant detection of
  X-ray Polarization in Sco X-1 with PolarLight and constraints on the corona
  geometry},'' {\em \apjl}, vol.~924, p.~L13, Jan. 2022.

\bibitem{Fomalont2001}
E.~B. {Fomalont}, B.~J. {Geldzahler}, and C.~F. {Bradshaw}, ``{Scorpius X-1:
  The Evolution and Nature of the Twin Compact Radio Lobes},'' {\em \apj},
  vol.~558, pp.~283--301, Sept. 2001.

\bibitem{Revnivtsev2006}
M.~G. {Revnivtsev} and M.~R. {Gilfanov}, ``{Boundary layer emission and Z-track
  in the color-color diagram of luminous LMXBs},'' {\em \aap}, vol.~453,
  pp.~253--259, July 2006.

\bibitem{Dai2007}
A.~{D'A{\'\i}}, P.~{{\.Z}ycki}, T.~{Di Salvo}, {\em et~al.}, ``{Broadband
  Spectral Evolution of Scorpius X-1 along Its Color-Color Diagram},'' {\em
  \apj}, vol.~667, pp.~411--426, Sept. 2007.

\bibitem{Titarchuk2014}
L.~{Titarchuk}, E.~{Seifina}, and C.~{Shrader}, ``{X-Ray Spectral and Timing
  Behavior of Scorpius X-1. Spectral Hardening during the Flaring Branch},''
  {\em \apj}, vol.~789, p.~98, July 2014.

\bibitem{Barnard2003}
R.~{Barnard}, M.~J. {Church}, and M.~{Ba{\l}uci{\'n}ska-Church}, ``{Physical
  changes during Z-track movement in Sco X-1 on the flaring branch},'' {\em
  \aap}, vol.~405, pp.~237--247, July 2003.

\bibitem{Church2012}
M.~J. {Church}, A.~{Gibiec}, M.~{Ba{\l}uci{\'n}ska-Church}, and N.~K.
  {Jackson}, ``{Spectral investigations of the nature of the Scorpius X-1 like
  sources},'' {\em \aap}, vol.~546, p.~A35, Oct. 2012.

\bibitem{Kong2021}
L.~D. {Kong}, S.~{Zhang}, L.~{Ji}, {\em et~al.}, ``{Luminosity Dependence of
  the Cyclotron Line Energy in 1A 0535+262 Observed by Insight-HXMT during the
  2020 Giant Outburst},'' {\em \apjl}, vol.~917, p.~L38, Aug. 2021.

\end{thebibliography}

\end{document}